\documentclass{elsart}
\input epsf.sty
\usepackage{natbib}
\usepackage{graphicx}
\usepackage{txfonts}
\input amssym.tex
\begin{document}
\runauthor{Ballero et al.}
\begin{frontmatter}

\title{The effects of Population III stars and variable IMF on the
  chemical evolution of the Galaxy}

\author[dip]{Silvia Kuna Ballero},
\author[dip]{Francesca Matteucci}
\author[oat]{ and Cristina Chiappini}
\address[dip]{Dipartimento di Astronomia, Universit\`a di Trieste, 
Via G.B. Tiepolo 11, 34124 Trieste, Italy}
\address[oat]{Osservatorio Astronomico di Trieste, 
Via G. B. Tiepolo 11,
34124 Trieste, Italy}



\begin{abstract}
 We have studied the effects of a hypothetical initial
  stellar generation  (Population III) containing only massive ($M > 10
  M_{\odot}$) and very massive stars ($M > 100 M_{\odot}$,
  Pair-Creation Supernovae) on the chemical evolution of the Milky Way. To
  this purpose, we have adopted a chemical evolution model 
  - the two-infall model from Chiappini et al. (1997) - which
  successfully reproduces the main observational features of the 
  Galaxy. Several sets of yields for very massive zero-metallicity
  stars have been tested: these stars in fact produce quite different 
  amounts of heavy elements, in particular $\alpha$-elements and iron,
  than lower mass stars. We have focused our attention on the chemical
  evolution of $\alpha$-elements, carbon, nitrogen and iron. It was 
  found that the effects of Population III stars on the Galactic
  evolution of these elements is negligible if only one or two
  generations of such stars occurred, whereas they produce quite
  different results from the standard models if they continuously
  formed for a longer period. Also the effects of a more strongly
  variable IMF were discussed and to this purpose we have made use of
  suggestions appeared in the literature to explain the lack of 
  metal-poor stars in the Galactic halo with respect to model
  predictions. In these cases the predicted variations in the
  abundance ratios, the SN rates and the G-dwarf metallicity
  distribution are more dramatic and always in contrast with
  observations, so we have concluded that a constant or slightly
  varying IMF remains the best solution.
  Our main conclusion is that if very massive stars ever existed
  they must  have formed only for a very short period of time (until
  the halo   gas reached the suggested threshold metallicity of
  $10^{-4}Z_{\odot}$  for the formation of very massive objects); 
  in this case, their effects on the evolution of the elements studied
  here was negligible also in the early halo phases. In other words,
  we cannot prove or disprove the existence of such stars on the basis
  of the available data on very metal poor stars.  Because of their
  large metal production and short lifetimes very massive primordial stars 
  should have enriched the halo gas to the metallicity of the most metal 
  poor stars known ([Fe/H] $\sim -5.4$) and beyond in only a few
  million years. This fact imposes constraints on the number of
  Pair-Creation Supernovae: we find that a number from 2 to 20 of such
  SNe occurred in our Galaxy depending on the assumed stellar
  yields.
 \end{abstract}

\end{frontmatter}


\section{Introduction}

There are several theoretical and observational reasons for suggesting
that the first stars (``Population III'' stars, hereafter PopIII),
which formed out of gas of primordial chemical composition, were more
massive than today. We recall that the Jeans mass, i.e., the minimum
mass necessary for a gas cloud to collapse under its self-gravity and
give rise to a protostar, scales as $T^{3/2}$, its effective
dependence on temperature being steeper if turbulent phenomena are
taken into account (Padoan et al., 1997). Metals are the most
effective coolants in gas clouds and they allow for the formation of
small mass protostars; in their absence, the cooling of the first
stars was based on the rotational-vibrational transitions of molecular
hydrogen. Hydrodynamical simulations of the collapse and fragmentation
of the  primordial gas clouds suggest that the very first stars should
have had masses larger than $100 M_{\odot}$ according to Bromm
et al. (2004), whereas Abel et al. (2002) give a mass range of
$30-300 M_{\odot}$. This high-mass biased star formation must have
lasted until a  certain critical metallicity $Z_{\mbox{PopIII}}$ was
reached, at which point star formation switched to a low-mass
dominated normal regime (e.g. a Scalo IMF). The latest calculations
constrain this value of $Z_{\mbox{PopIII}}$ in the range
$(10^{-6}-10^{-4})\cdot Z_{\odot}$ (Schneider et al., 2002).
The IMF may also depend on the ionization state of the gas and
on the background of ionizing photons which could inhibit the
formation of H$_2$ and require even bigger halos for the creation of
stars (see e.g. Haiman et al., 2000).

Being so massive, a PopIII star should start its main sequence with
CNO burning, but since these nucleids were not present in the
primordial gas, the star is forced to burn hydrogen via the $p-p$ 
chain, which is much less effective in counterbalancing the effect of
gravity; as a result, the star contracts more with respect to a normal
PopI/PopII star, until it makes a trace of $^{12}$C by 3-$\alpha$
reaction, at which point it burns H via the CNO cycle. Such a more
compact star evolves at a higher surface temperature, thus producing
a great deal of ionizing photons. The compactness is also caused by
the lower opacity due to the absence of metals. 
Owing to the lack of metals in their envelope, PopIII stars suffer
from less mass loss during their lifetime if compared to massive PopII
and PopI stars. This gives way, at the end of their lives, to a
more probable formation of massive black holes and the so-called 
Pair-Creation Supernovae \mbox{(PCSNe)}, which leave no remnant after
their explosion (Rakavy \& Shaviv, 1967; Ober et al. 1982).

From an observational point of view, there are several cosmological
and astrophysical problems which could be solved if one or more
generations of massive metal-free stars existed. Recently, the WMAP
experiment, by observing the large-angle polarization of the CMB, has
provided constraints on the number of ionizing photons produced from
either the first massive stars or AGNs at $z \ge 15$ (e.g. Cen, 2003). 
Furthermore, the Lyman-$\alpha$ forest at $z\simeq4-5$, as seen in the
QSO light looks considerably metal-polluted (Schaye et al. 2003),
suggesting an early stellar population which enriched the
  intergalactic medium via strong winds.

The problems related to galactic chemical evolution which could be 
solved by the introduction of PopIII are the pre-enrichment and the
necessity of primary nitrogen in the early phases of the Galaxy. In
fact, no zero metal stars were ever observed in the Galactic halo: the
most metal poor star discovered so far (Frebel et al., 2005) has
$[$Fe/H$] = -5.4 \pm 0.2$ . This may indicate that the first generation of
stars in our Galaxy, which enriched the gas out of which were born the
Population II stars we observe today, were massive and died soon
(even though it cannot be excluded that they were rare enough
  to be inadequately sampled by surveys so far). If this is the case, we
could infer important constraints on their nucleosynthetic properties
from the chemical abundances of the most metal-poor stars. As to N
production, it is usually assumed that in massive stars nitrogen is a
secondary element, i.e., it is created from the seed nuclei of carbon
during the CNO cycle; this implies that the abundance of nitrogen
should decrease considerably towards lower metallicities. This is at
variance with the most recent data of Spite et al. (2005) for N
abundances in the Galactic halo, which show a flat [N/Fe] for [Fe/H]$<
-2.5$. In this respect, PopIII stars could be good candidates for
primary production of N from massive stars (see also Marigo et al.,
2003; Meynet et al, 2003; Heger et al. 2000).

As far as $\alpha$/Fe ratios in very metal poor stars are concerned
there is apparently no need to invoke PopIII star nucleosynthesis
(e.g. Fran\c cois et al. 2004), but since PopIII star nucleosynthesis
has been invoked to explain some peculiar behaviours of Fe-peak
elements (Mn, Co, Cr, Ni, Zn) (see Umeda \& Nomoto 2002, 2005) it is
interesting to check the effect of PopIII stars also on 
$\alpha$-elements and Fe. In particular, we aim at testing whether, in
spite of the different $\alpha$/Fe ratios predicted for PopIII stars
relative to normal stars, the inclusion of these very massive objects
in chemical evolution models can still be acceptable. In other words,
we aim at investigating if the evolution in time of Galactic chemical
abundances can impose constraints on the characteristics and duration
of the PopIII phase. Thus we have incorporated the nucleosynthetic
products from PopIII stars of $\alpha$ elements, C, N and Fe in a
detailed chemical evolution model. 
To our knowledge this is the first test of this kind since the
previous ones had only compared the observed abundance ratios in metal
poor stars with the stellar production ratios.

In \S 2 we describe this model; in \S 3 we discuss the way we
have modelled the PopIII phase; in \S 4 we show the results; in \S 5
we briefly examine the effects of a variable IMF similar to that
proposed by Larson (1998) (namely, ``top-heavy'' at the earliest times) and
in \S 6 we draw conclusions.

\section{The basic model}
\label{sec:mod}

The model for the chemical evolution of the Galaxy on which we based
our investigation is essentially the two-infall model developed by
Chiappini et al. (1997), in which the two main components of the
Galaxy (halo/thick disk and thin disk) are supposed to be formed by
two separate 
episodes of accretion of extragalactic gas of primordial chemical
composition. During the first episode, the halo is formed on a
timescale of 1 Gyr; afterwards, the thin disk forms, with a timescale which
is proportional to the Galactocentric distance (``inside-out''
formation) such that the timescale for the solar neighbourhood is
about 7 Gyr (much longer than for the halo).

It is worth noting that the present model does not refer to
  the hierarchical structure formation scenario, but it makes use of a
  backward approach, in the sense that we start from the present-time
  properties and try to derive the initial conditions which will lead
  to such properties. This approach is not in contradiction with the
  cosmological approach which starts from precise initial conditions,
  but is rather complementary. A comparison of the ``forward'' and
  ``backward'' scenarios can be, in fact, very useful. The reason why
  we do not adopt the forward approach here is that it still lacks
  sufficient numerical resolution to resolve single galaxies and
  contains large uncertainties on basic physical processes. However,
  we stress that computing galaxy evolution at high resolution and in
  a cosmological context will ultimately be successful.
  Recently, Sommer-Larsen et  al. (2003) performed simulations of
  galaxy formation in a standard $\Lambda$CDM cosmology, including
  star formation and feedback effects and found that the gas
  infall rate onto the solar cylinder is exponentially declining with
  time, with an infall timescale comparable to that used in this
  model and obtained with the backward approach. Moreover, the
  predictive power of models of chemical evolution, regarding
  abundances, gas fraction, supernova rates etc., is still greater
  than that of models taking into account the hierarchical clustering
  paradigm.

The star formation rate is parametrized as follows:
\begin{equation}
\psi(r,t) = 
\left\{
\begin{array}{ll}
\tilde{\nu}\sigma_{tot}^{k_1}(r,t)\sigma_{g}^{k_2}(r,t) 
  & \mbox{ if } \sigma_g(r,t)  >   \sigma_{thr}\\
0 & \mbox{ if } \sigma_g(r,t) \leq \sigma_{thr}
\end{array}
\right.
\end{equation}
where $k_1 = 0.5$ and $k_2 = 1.5$; $\tilde{\nu} = 1$ Gyr$^{-1}$ is the
star formation efficiency, $\sigma_g$ and $\sigma_{tot}$ are
respectively the surface gas density and the total surface mass
density (gas + stars + remnants), $\sigma_{thr} = 7$
$M_{\odot}$pc$^{-2}$ is a threshold gas 
density below which star formation stops (Kennicutt, 1989).

The initial mass function (IMF) has the following functional form:
\begin{equation}
\varphi(m) \propto m^{-(1+x)} \;\;\; x =
\left\{
\begin{array}{lll}
1.35 & \mbox{if}& 0.1 \leq m/M_{\odot} \leq 0.6 \\
1.7 & \mbox{if} & 0.6 \leq m/M_{\odot} \leq 80
\end{array}
\right.
\label{eqimf}
\end{equation}
which closely resembles the one proposed, on
observational grounds, by Scalo (1986), except that it has two slopes
instead of three. 
Stars do not form outside the mass range \mbox{$0.1 \leq M/M_{\odot} \leq 80$.}

The stellar yields were adopted as follows (the choices
below correspond to model~5 of Chiappini et al. 2003a, CRM03):

\begin{itemize}

\item[-]\emph{Low and intermediate mass stars} ($0.8 \le M/M_{\odot}
  \le 8.0$): yields are taken from Van den Hoek \& Groenewegen (1997)
  as a function of initial metallicity. These stars produce He,
  $^{12}$C, N and heavy $s$-elements. 

\item[-]\emph{Massive stars} ($M \geq 10M_{\odot}$): these stars give
  rise to Type II SNe. Their chemical yields are those from Woosley \&
  Weaver (1995) for solar metallicity, extrapolated to the upper limit
  of the IMF\footnote{Woosley \& Weaver (1995) originally calculated
  stellar yields for masses in the range $12-40M_{\odot}$ and found
  that above this mass a considerable reimplosion may occur with
  severely reduced yields and likely black hole formation. However, it
  was shown by Fran\c cois et al. (2004) that if black hole formation
  occurs in this mass range it is impossible to fit the data at low
  metallicity; therefore, since very few calculations are available in
  the mass range $40-100M_{\odot}$ one has to perform an extrapolation. 
  The models of Woosley \& Weaver (1995) do not include mass loss,
  which is an acceptable assumption for Pop III stars but could be
  less realistic for later generations.}; they contribute to the
  enrichment in $\alpha$-elements, some Fe, $r$-elements and light
  $s$-elements.

\item[-]\emph{Type Ia supernovae}: bearing in mind the W7 (single
  degenerate) model from Nomoto et al. (1984), yields are taken from
  Iwamoto et al. (1999) which is an updated version of model
  W7. These supernovae produce most Fe and traces of light elements.

\item[-]\emph{Novae}: nucleosynthesis from nova outbursts is
  included (Jos\'e \& Hernanz, 1998); novae can contribute to the
  production of CNO isotopes and Li (Romano \& Matteucci, 2003).

\end{itemize}

This model reproduces well the observational constraints for the
solar neighborhood, if we exclude N, overproduced by the yields of Van
den Hoek and Groenewegen (1997) for low and intermediate mass stars,
and Mg, underproduced by the yields of Woosley and Weaver (1995) for
massive stars. The model also reproduces well the chemical
evolution of abundance ratios as a function of [Fe/H] for all
elements, with the exception of nitrogen (see Fran\c cois et
al. 2004); in fact, since the production of nitrogen from massive stars
is assumed to be secondary, that model predicts a steep decrease of $[$N/Fe$]$
vs. $[$Fe/H$]$ towards lower values of [Fe/H], at variance with the new
data from Spite et al. (2005) which point towards a roughly constant and solar
value for $[$N/Fe$]$ at very low iron abundances.

\section{Modelling Population III}

Modelling the PopIII phase requires the choice of a critical
metallicity $Z_{\mbox{PopIII}}$,  below which the IMF and the stellar
yields are characteristic of metal-free stars. To do that, we need a
functional form and a mass range for the IMF and the adoption of zero
metallicity stellar yields. 

Concerning the first point, we ran models with $Z_{\mbox{PopIII}}$ in
the range $10^{-6}Z_{\odot}-0.1Z_{\odot}$ (the upper limit of this
range is not realistic and has a merely investigative
meaning). However, we are showing the results only for
$Z_{\mbox{PopIII}} \geq 10^{-4}Z_{\odot}$, since we saw no differences
in the models with $Z_{\mbox{PopIII}}$ equal to $10^{-6}$, $10^{-5}$
and $10^{-4} Z_{\odot}$. This can be understood if we bear in mind
that it takes only a few supernovae to enrich the interstellar
medium significantly, and as a matter of fact all of these critical
metallicities correspond to a unique generation of PopIII stars.

We also ran models with a variety of mass ranges and IMFs for the
PopIII phase, but hereafter we will present only models where low and
intermediate mass stars were not included in the primordial IMF. This
primordial IMF has the same form as in Eq. \ref{eqimf}, except that, the
turning point being outside the considered mass range, there is a
unique slope. In other words:
\begin{equation}
\varphi(m)\propto m^{-(1+x)}, \;\; M_L\leq M \leq M_U
\label{eq:primf}
\end{equation}
where $x=1.7$, $M_L=10M_{\odot}$ and $M_U$ depends on the considered
nucleosynthesis, as we shall see next.
When $Z_{\mbox{PopIII}}$ is reached, the IMF switches to the one shown in
Eq. \ref{eqimf}.

\subsection{Nucleosynthesis}

The chemical evolution and nucleosynthesis of zero-metal massive stars
have been computed since the early eighties (Ober et al., 1983, El Eid
et al., 1983) and the work in this field has continued until very
recently (Woosley \& Weaver, 1995; Heger \& Woosley, 2002; Umeda \&
Nomoto, 2002; Heger et al., 2003; Chieffi \& Limongi, 2002 and
2004). Also yields for low (D'Antona, 1982) and intermediate
(e.g. Siess et al., 2002) mass stars are available, but we are not
using them right now since we are only considering massive stars for
the primordial IMF. We supposed that zero metallicity yields hold
until $Z_{\mbox{PopIII}}$ is reached, then the nucleosynthesis becomes
the standard one.  
For our purpose, we have considered four sets of nucleosynthetic yields:

\begin{itemize}
\item[-]\emph{Ober et al., 1983}: these authors computed the 
yields for zero metallicity stars with main sequence masses in the range 
$80-500M_{\odot}$. They predicted that these objects should produce mainly 
oxygen and strongly underproduce nitrogen and Fe-peak nuclei,
supposedly because they were lacking the appropriate reaction
  network in their code to produce iron. They also predicted a
substantial mass loss during the main sequence phase, in contrast with
all the other considered sets of yields. 
These calculations are old and likely to have been superceded by the
most recent ones; moreover, the lack of iron in their yields leads to
a severe overestimation of the $[$el/Fe$]$ ratios at low values of [Fe/H]
with respect to observations for $\alpha$-elements and C (see
Fig. \ref{fig 1}), even for the lowest values of $Z_{\mbox{PopIII}}$.

\begin{figure}
\centering
\includegraphics[width=.84\textwidth,clip,trim=0 290 0 0]{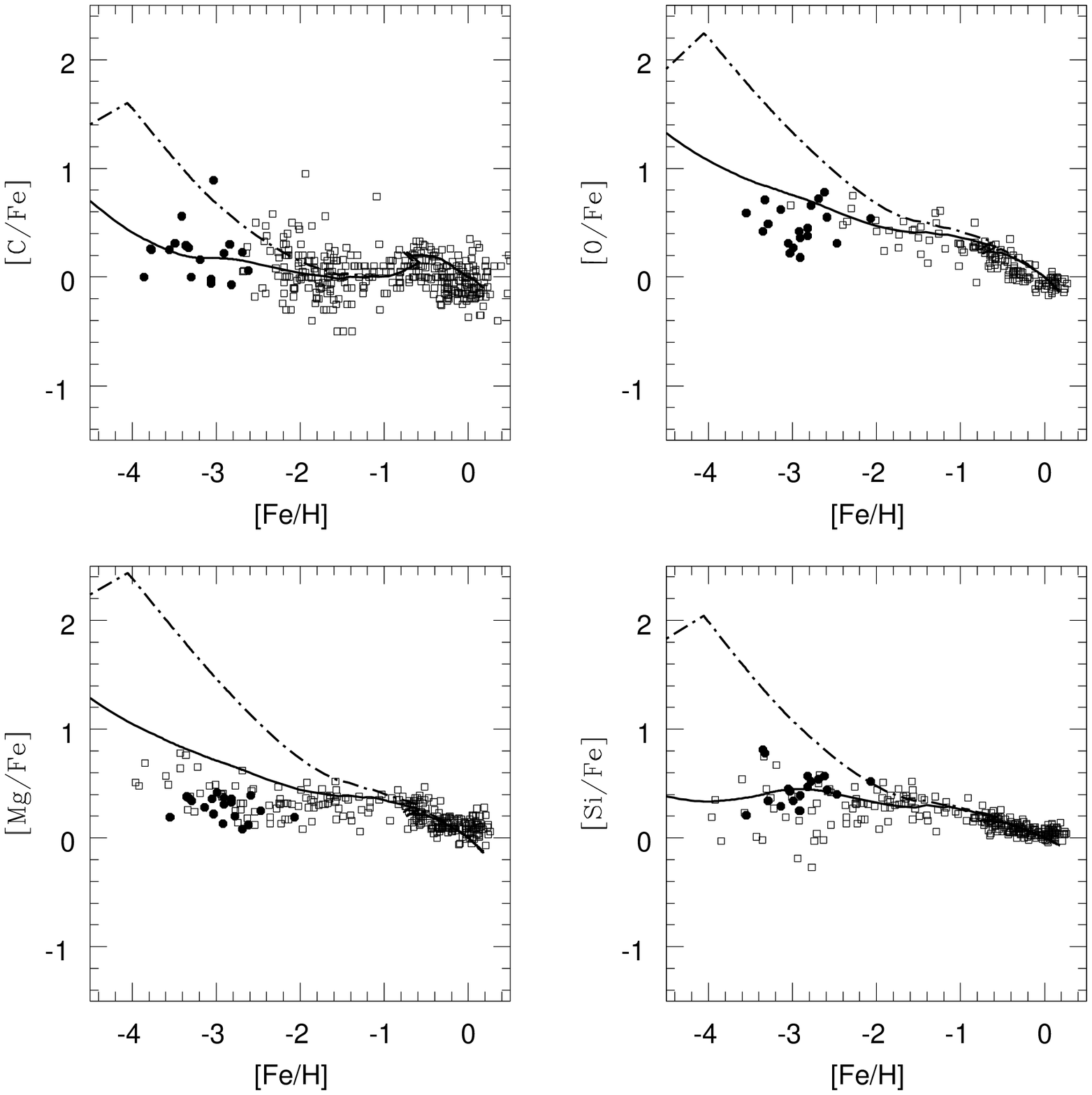}
\caption{Evolution of abundance ratios [C/Fe] and [O/Fe] with
  metallicity  with yields of Ober et al. (1983), \mbox{$Z_{\mbox{PopIII}} =
  10^{-6}Z_{\odot}$} (dashed-dotted line), compared to the results of
  CRM03 model (their model 5, solid line), both normalized to the solar
  values. One can notice the strong overproduction of all the elements
  with respect to iron at low [Fe/H]. Data are taken from Chiappini
  et al. (1999, and references therein) for the disk phase; for the
  halo phase, data are from Spite et al. (2005) for C and from Cayrel
  et al. (2004) for O.}
\label{fig 1}
\end{figure}

\item[-]\emph{Woosley \& Weaver, 1995 + Heger \& Woosley, 2002}
  (hereafter HW): we have combined two sets of yields, the first one
  (Woosley \& Weaver, 1995, their case A for $M=12-22M_{\odot}$, case B
  for $M=25-30M_{\odot}$, case C for $M = 35-40M_{\odot}$, in order to
  achieve the maximum enrichment) for stars with main sequence masses
  in the range  $12-40 M_{\odot}$, the second one (Heger \&
  Woosley, 2002) for stars with He-core masses in the range
  $64-133M_{\odot}$, corresponding to main sequence masses of
  $140-260M_{\odot}$, which explode as PCSNe leaving no remnant
  (thus, referring to Eq. \ref{eq:primf}, $M_U=260M_{\odot}$). For
  He-core masses larger than 133$M_{\odot}$, these authors found that
  a black hole is formed and no nucleosynthesis is produced. They also
  found that the same situation is likely to occur in zero-metal stars
  with main sequence masses in the range $40-140M_{\odot}$;
  however, we also computed models where black hole formation is not
  considered in this mass range, performing linear interpolation
  between 40 and $140M_{\odot}$ instead \footnote{Again,
  linear interpolation may not be the best choice, but due to the lack
  of available yields in this mass range it would otherwise be
  impossible to investigate what would be the effect of a
  non-negligigle chemical enrichment from these stars.}. Pre-SN
  evolution is calculated at constant  mass, so mass loss yields are
  not taken into account. 
\item[-]\emph{Umeda \& Nomoto, 2002} (hereafter UN): these authors
  calculated the evolution and nucleosynthesis of PopIII stars
  for the main sequence mass ranges $13-30M_{\odot}$ and
  $150-270M_{\odot}$; the latter includes PCSN progenitors. They
  compared the abundance ratios of halo stars with the production
  ratios in PCSNe and concluded that they are not in agreement;
  however, since this result is based on a simple approach, we did not
  take it for granted and computed both models where
  PCSNe are considered ($M_U = 270M_{\odot}$), and models where they
  are not ($M_U = 80M_{\odot}$). Referring to their paper, we adopted
  the yields for $E_{51}=1$. The authors suppose that black holes form
  in the mass  range $30-150M_{\odot}$, but do not rule out a
  supernova explosion, releasing some small quantity of elements,
  accompanying black hole formation. Given these uncertainities, we
  still computed models both with and without black hole formation. In
  this case as well, mass loss is not taken into account.
  We notice that the UN yields, contrarily to HW yields, are
  rather sparse in mass and therefore their incorporation in chemical
  evolution studies might lead to larger uncertainties. 
\item[-]\emph{Chieffi \& Limongi, 2002-2004} (hereafter CL): this set
  was also obtained by combining the results of two papers, the first
  one (Chieffi \& Limongi, 2004) for masses in the range
  $10-35M_{\odot}$ and the second one (Chieffi \& Limongi, 2002) for
  stars with masses equal to 50 and $80M_{\odot}$. Since these are the
  only yields computed by these authors, we restricted the mass range
  to $10-80M_{\odot}$, so that no PCSN scenario occurs; in addition,
  though not explicitly stated by the authors, neither black hole 
  formation nor mass loss are considered. These yields are peculiar in
  the fact that they adopt a particularly low rate for the reaction
  $^{12}C(\alpha,\gamma)^{16}O$, which leads to an overproduction of C
  with respect to other yields. The authors made this choice as a
  result of a comparison with the chemical composition of an
  ``average'' star, representing the halo stars with $[Fe/H] \le -3.3$
  (under the hypothesis that they are enriched by a few supernova
  explosions from the first stars) and bearing in mind that the cross
  section for that reaction is very poorly constrained. In order to
  obtain the maximum enrichment, we chose the yields calculated with
  the lowest mass-cut. 
\end{itemize}
Table \ref{tab-mod1} summarizes the features of the models we are
presenting and discussing here. In Column 1 models are
identified as follows: the first letter refers to the adopted yields,
the number is associated with $Z_{\mbox{PopIII}}$ (e.g. ``1'' stands for
$Z_{\mbox{PopIII}} = 10^{-1}Z_{\odot}$), the following letter
states the mass range if more than one mass range is considered
for a given nucleosynthesis (``P'' means that PCSNe are included, ``M'' that
they are not); a final ``b'' means that black hole formation is assumed
in a certain mass range which is specified in Column 4. In Column 2 we
show the yield source for PopIII stars; in Column 3 we specify the
assumed mass range for PopIII stars; finally, in Column 5 we show
the adopted threshold metallicity.
\begin{table*}
\centering
\begin{tabular}{lllll}
\hline
\hline
Model name &Yields &Mass range ($M_{\odot}$) & BH range ($M_{\odot}$) &
$Z_{\mbox{PopIII}}$ ($Z_{\odot}$)\\ 
\hline
\hline
H1& HW& $10-270 $ & none & $0.1 $ \\
H2& HW&$10-270 $ & none &$10^{-2} $ \\
H3& HW&$10-270 $ & none &$10^{-3} $ \\
H4& HW&$10-270 $ & none &$10^{-4} $ \\
\hline
H1b& HW& $10-270 $ & $40-140 $ &$0.1 $ \\
H2b& HW& $10-270 $ & $40-140 $ &$10^{-2} $ \\
H3b& HW& $10-270 $ & $40-140 $ &$10^{-3} $ \\
H4b& HW& $10-270 $ & $40-140 $ &$10^{-4} $ \\
\hline
U1P&UN& $10-270 $ & none &$0.1 $ \\
U2P&UN& $10-270 $ & none &$10^{-2} $ \\
U3P&UN& $10-270 $ & none &$10^{-3} $ \\
U4P&UN& $10-270 $ & none &$10^{-4} $ \\
\hline
U1Pb&UN& $10-270 $ & $30-150 $ &$0.1 $ \\
U2Pb&UN& $10-270 $ & $30-150 $ &$10^{-2} $ \\
U3Pb&UN& $10-270 $ & $30-150 $ &$10^{-3} $ \\
U4Pb&UN& $10-270 $ & $30-150 $ &$10^{-4} $ \\
\hline
U1M&UN& $10-80 $ & none &$0.1 $ \\
U2M&UN& $10-80 $ & none &$10^{-2} $ \\
U3M&UN& $10-80 $ & none &$10^{-3} $ \\
U4M&UN& $10-80 $ & none &$10^{-4} $ \\
\hline
U1Mb&UN& $10-80 $ &$30-150 $ & $0.1 $ \\
U2Mb&UN& $10-80 $ &$30-150 $ & $10^{-2} $ \\
U3Mb&UN& $10-80 $ &$30-150 $ & $10^{-3} $ \\
U4Mb&UN& $10-80 $ &$30-150 $ & $10^{-4} $ \\
\hline
C1& CL&$10-80 $ & none & $0.1 $ \\
C2& CL&$10-80 $ & none &$10^{-2} $ \\
C3& CL&$10-80 $ & none &$10^{-3} $ \\
C4& CL&$10-80 $ & none &$10^{-4} $ \\
\hline
\hline
\end{tabular}
\vspace{6pt}
\caption[]{Definition of PopIII models: the first column specifies the
  model names; the second shows the adopted nucleosynthesis;
  the third and fourth columns show respectively the considered mass
  range and the black hole formation range; the last column
  specifies the critical metallicity marking the end of PopIII phase.}
\label{tab-mod1}
\end{table*}

\section{Results}

\subsection{Models with PCSNe}

We first present the results of all those models which
include PCSNe in their primordial IMF; these are all
the models with yields of HW (from H1 to H4b in Table
\ref{tab-mod1}), for which $M_U=260M_{\odot}$, and half of the models
with yields of UN (from U1P to U4Pb in Table \ref{tab-mod1}) for which
$M_U=270M_{\odot}$.

In Table \ref{tab-res1} are shown the solar abundances by mass (i.e.,
abundances predicted for the gas $4.5$ Gyr ago), compared to the same
results from our standard model (similar to CRM03 - see the tables for
details).


\begin{table*}
\centering
\caption[]{Solar abundances by mass predicted by PopIII models with
  PCSNe and yields from Heger \& Woosley (2002) and Woosley \& Weaver
  (1995); data for solar abundances are from Asplund et
  al. (2004). $\Delta  t_{PopIII}$ is the time duration of the PopIII
  phase (from the  beginning of star formation) and SN\# is 
  the estimated number of supernovae which exploded during such phase.}
\label{tab-res1}
\vspace{6pt}
\begin{center}
\begin{tabular}{lcccccc}
\hline
\hline$X_i$ & Standard & H1& H2& H3& H4& Obs.\\
\hline
C& 1.6E$-3$&1.6E$-3$&1.7E$-3$&1.7E$-3$&1.7E$-3$&2.1E$-3$  \\
N& 9.6E$-4$&9.6E$-4$&9.6E$-4$&9.6E$-4$&9.6E$-4$&5.9E$-4$  \\
O& 5.3E$-3$&5.3E$-3$&5.3E$-3$&5.3E$-3$&5.3E$-3$&5.1E$-3$  \\
Mg& 2.5E$-4$&2.3E$-4$&2.3E$-4$&2.3E$-4$&2.3E$-4$&5.7E$-4$ \\
Si& 8.3E$-4$&5.3E$-4$&5.3E$-4$&5.3E$-4$&5.3E$-4$&6.3E$-4$  \\
Fe& 1.1E$-3$&1.1E$-3$&1.1E$-3$&1.1E$-3$&1.1E$-3$&1.1E$-3$ \\
$Z_{\odot}$& 1.2E$-2$& 1.2E$-2$& 1.2E$-2$& 1.2E$-2$& 1.2E$-2$& 1.2E$-2$  \\
\hline
$\Delta t_{PopIII}$ ($Myr$)&-&10&4&3&2&-\\
SN\# & - & 2500 & 80 & 8 & 3 & -\\
\hline
\hline
\end{tabular}
\end{center}

\centering
\begin{center}
\begin{tabular}{lcccccc}
$X_i$ & Standard & H1b& H2b& H3b& H4b& Obs.\\
\hline
C& 1.6E$-3$&1.7E$-3$&1.7E$-3$&1.7E$-3$&1.7E$-3$&2.1E$-3$ \\
N& 9.6E$-4$&9.6E$-4$&9.6E$-4$&9.6E$-4$&9.6E$-4$&5.9E$-4$ \\
O& 5.3E$-3$&5.3E$-3$&5.3E$-3$&5.3E$-3$&5.3E$-3$&5.1E$-3$ \\
Mg& 2.5E$-4$&1.7E$-4$&1.7E$-4$&1.7E$-4$&1.7E$-4$&5.7E$-4$ \\
Si& 8.3E$-4$&4.9E$-4$&4.9E$-4$&4.9E$-4$&4.9E$-4$&6.3E$-4$ \\
Fe& 1.1E$-3$&1.1E$-3$&1.1E$-3$&1.1E$-3$&1.1E$-3$&1.1E$-3$\\
$Z_{\odot}$& 1.2E$-2$& 1.1E$-2$& 1.1E$-2$& 1.1E$-2$& 1.1E$-2$& 1.2E$-2$  \\
\hline
$\Delta t_{PopIII}$ ($Myr$) &-&17&5&3&2&-\\
SN\# & - & 5500 & 200 & 15 & 2 & - \\
\hline
\hline
\end{tabular}
\end{center}
\end{table*}
\begin{table*}
\caption{Same as table \ref{tab-res1} but with yields of Umeda \&
  Nomoto (2002).}
\label{tab-res3}
\vspace{6pt}
\centering
\begin{center}
\begin{tabular}{lcccccc}
\hline
\hline
$X_i$ & Standard & U1P& U2P& U3P& U4P& Obs.     \\
\hline
C& 1.6E$-3$&1.6E$-3$&1.7E$-3$&1.7E$-3$&1.7E$-3$&2.1E$-3$\\
N& 9.6E$-4$&9.6E$-4$&9.6E$-4$&9.6E$-4$&9.6E$-4$&5.9E$-4$\\
O& 5.3E$-3$&5.3E$-3$&5.3E$-3$&5.3E$-3$&5.3E$-3$&5.1E$-3$\\
Mg& 2.5E$-4$&3.8E$-4$&3.8E$-4$&3.8E$-4$&3.8E$-4$&5.7E$-4$\\
Si& 8.3E$-4$&8.2E$-4$&8.1E$-4$&8.1E$-4$&8.1E$-4$&6.3E$-4$\\
Fe& 1.1E$-3$&9.3E$-4$&9.3E$-4$&9.3E$-4$&9.3E$-4$&1.1E$-3$\\
$Z_{\odot}$& 1.2E$-2$& 1.2E$-2$& 1.2E$-2$& 1.2E$-2$& 1.2E$-2$& 1.2E$-2$\\
\hline
$\Delta t_{PopIII}$ ($Myr$)&-&10&4&3&2&-  \\
SN\# & - & 2000 & 80 & 8 & 3 & -  \\
\hline
\hline
\end{tabular}
\end{center}
\centering

\begin{center}

\begin{tabular}{lcccccc}
$X_i$ & Standard& U1Pb& U2Pb& U3Pb& U4Pb& Obs.\\
\hline
C& 1.6E$-3$&1.7E$-3$&1.7E$-3$&1.7E$-3$&1.7E$-3$&2.1E$-3$\\
N& 9.6E$-4$&9.6E$-4$&9.6E$-4$&9.6E$-4$&9.6E$-4$&5.9E$-4$\\
O& 5.3E$-3$&5.3E$-3$&5.3E$-3$&5.3E$-3$&5.3E$-3$&5.1E$-3$\\
Mg& 2.5E$-4$&2.1E$-4$&2.1E$-4$&2.1E$-4$&2.1E$-4$&5.7E$-4$\\
Si& 8.3E$-4$&5.7E$-4$&5.6E$-4$&5.6E$-4$&5.6E$-4$&6.3E$-4$\\
Fe& 1.1E$-3$&9.0E$-4$&9.0E$-4$&9.0E$-4$&9.0E$-4$&1.1E$-3$\\
$Z_{\odot}$& 1.2E$-2$& 1.1E$-2$& 1.1E$-2$& 1.1E$-2$& 1.1E$-2$& 1.2E$-2$\\
\hline
$\Delta t_{PopIII}$ ($Myr$)&-&19&6&3&2&-\\
SN\# & - & 5000 & 250 & 7 & 2 & - \\
\hline
\hline
\end{tabular}
\end{center}
\end{table*}


These tables
include the duration of the PopIII phase and the number of supernova
explosions which occurred before $Z_{\mbox{PopIII}}$ was reached in
the ISM. Several quantities, such as the star formation rate, the
surface gas and star density, the SN rate, etc., calculated
in the solar neighbourhood at the present time (i.e., the age of the
Galaxy $t_G=13.7$ Gyr) do not suffer from variations with respect to
our basic model for any adopted nucleosynthesis, therefore we will not
show the results concerning them. This fact is plausible since, as
shown in the tables, the PopIII phase is always extremely short (a few
Myr) even for the highest critical metallicities and, moreover, most
hypotheses which determine these values are left unchanged in our
models with respect to CRM03 model.

Our calculations of both the time duration of the PopIII phase
and the number of PCSNe which exploded during this phase are a
   first-approximation estimate and may not be very accurate, since
   the modeling of the first phases of Galactic collapse may be
   oversimplified, especially if we consider our adoption of the
   instantaneous mixing approximation. However, the time duration of
   the PopIII phase and the corresponding number of PCSNe cannot be
   much longer than this since it was shown (Oey, 2003) that even if
   the early evolution took place inhomogeneously (which still has to
   be verified via measurements of chemical abundances at very low
   metallicities), it rapidly became essentially homogeneous. In any
   case, our estimates can be regarded as order-of-magnitude estimates.

One can immediatly notice that substantial variations in the
predicted solar abundances by mass occur 
only for Fe, Mg and Si, and in any case the results are very similar
(within a factor of 2) to those of CRM03 model. These variations are
practically the same, within a given nucleosynthetic set, for every value
of the critical metallicity. Focusing on these elements, an
improvement of the agreement with observations would be achieved if,
with respect to our fiducial model, Mg were overproduced by a factor
of about two and Si were slightly underproduced, while for Fe there is
no need for variations. 
In general, models with the yields of HW tend to overproduce Fe and
underproduce Mg and Si, the latter effect being enhanced when black
hole formation is taken into account; however, the iron overproduction
may result from our adoption of those yields which led to the highest
enrichment. The PCSN yields of Fe are strongly dependent on the mass
cut and a different choice of this parameter could reduce the Fe yield
from this population. Instead, when we use the yields of UN, 
Fe and Si are underproduced, and Mg is overproduced with respect to
the model without PopIII, if black holes do not form; if they form,
all of the three elements are underproduced. Therefore, none of our
PopIII models with PCSNe gives a better agreement with observations if
compared to our basic standard model. Note that for those elements 
which are mainly produced by low and intermediate mass stars, which
contribute to the chemical enrichment only in the latest stages of
Galactic evolution, no variation is seen in the predicted solar
abundances with respect to our fiducial model.

\begin{figure*}
\centering
\includegraphics[width=.84\textwidth,clip,trim=0 290 0 0]{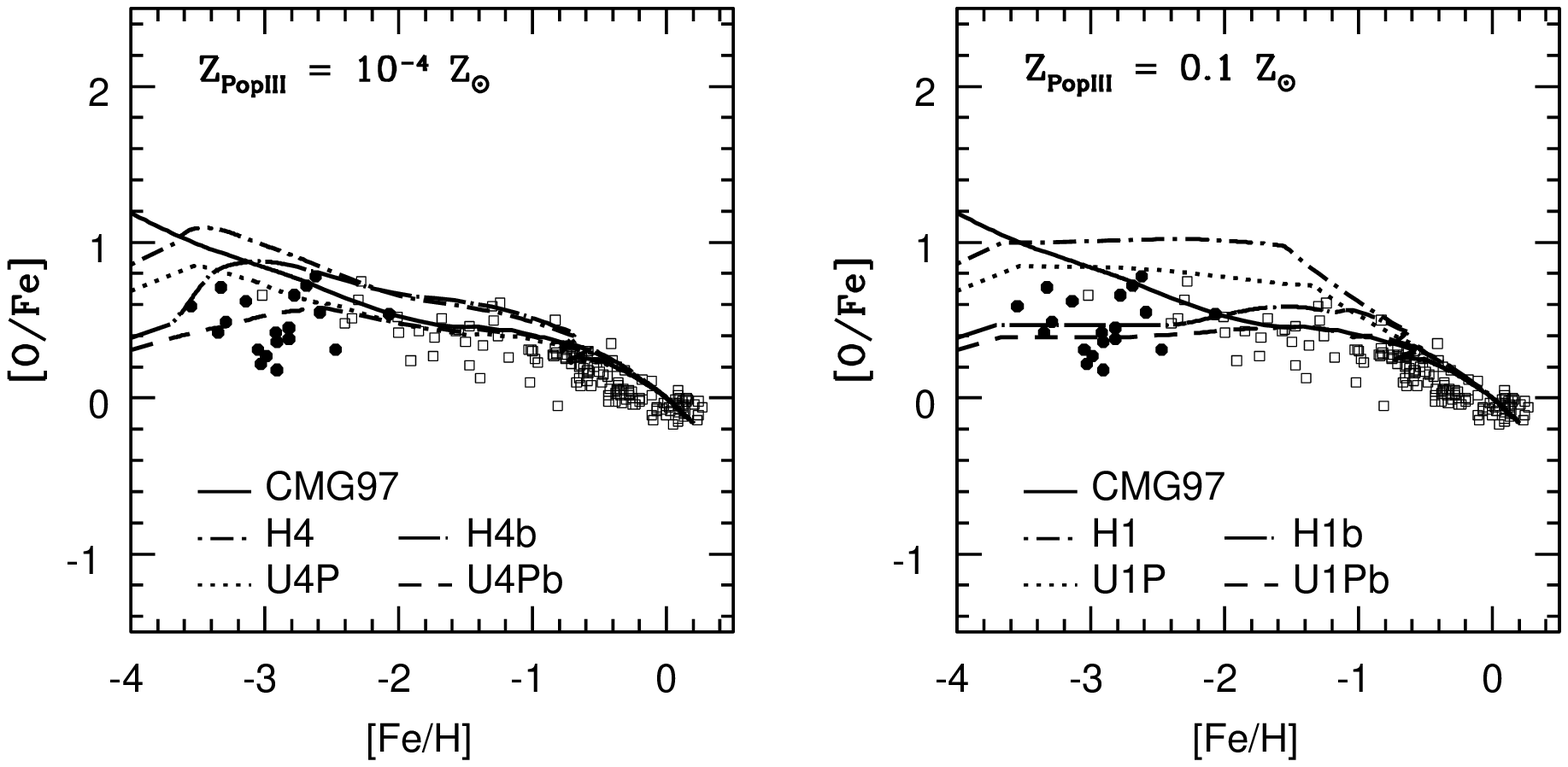}
\includegraphics[width=.84\textwidth,clip,trim=0 290 0 0]{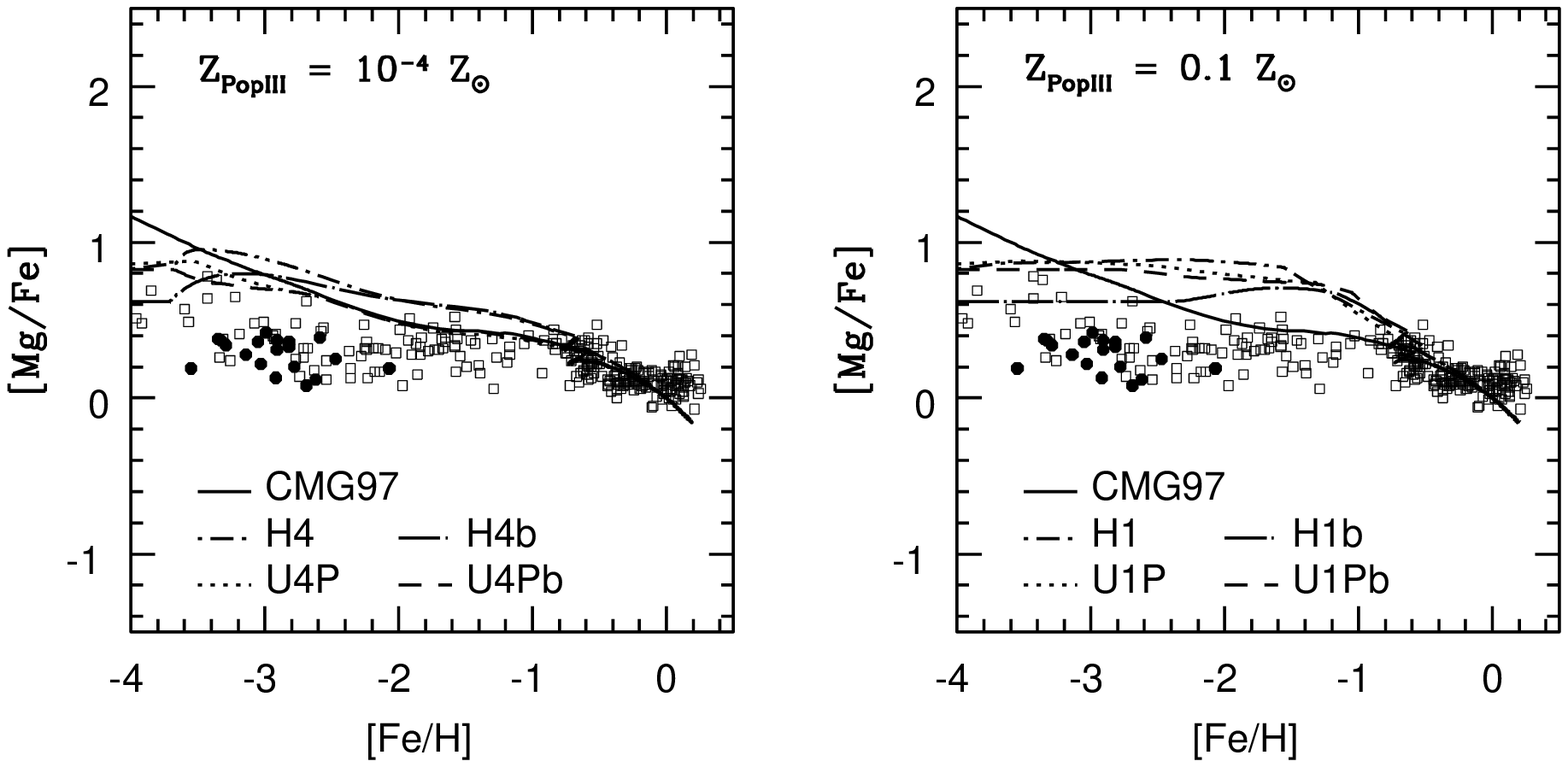}
\includegraphics[width=.84\textwidth,clip,trim=0 270 0 0]{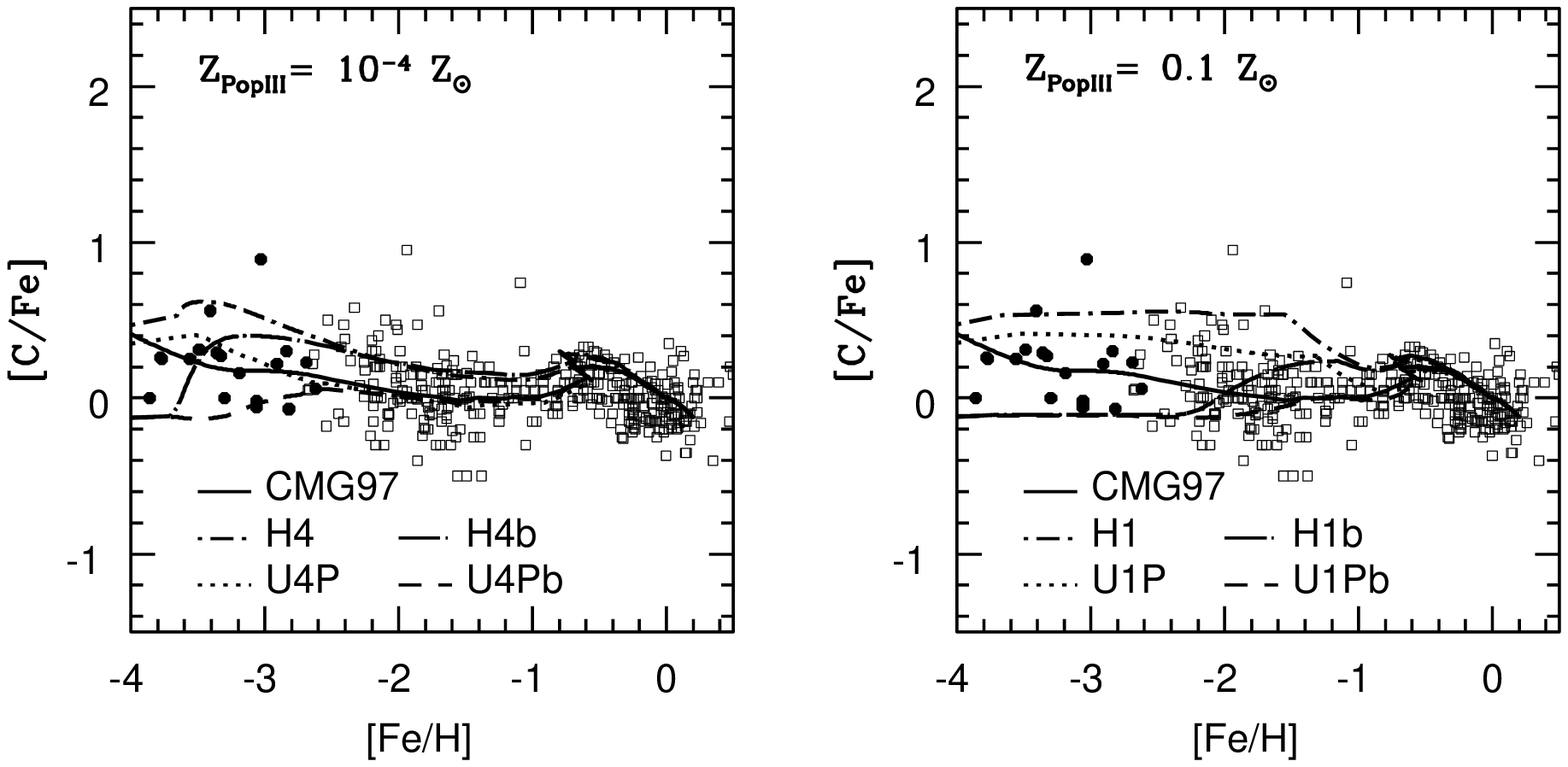}
\caption[]{Evolution of $[$O/Fe$]$, $[$Mg/Fe$]$ and $[$C/Fe$]$ with
  [Fe/H] in the model of CRM03 (solid line) and
  in the models with PopIII and PCSNe included in the IMF, for the
  lowest and highest
  considered values of the critical metallicity. Data are from
  Chiappini et al. (1999, and references therein) for the disk phase
  (open squares); for the halo phase (filled circles) data are from
  Cayrel et al. (2004) for O and Mg and from Spite et al. (2005) for C.}
\label{fig 2}
\end{figure*}

Fig. \ref{fig 2} shows the evolution with [Fe/H] of the abundance
ratios $[$X/Fe$]$ for X corresponding to two $\alpha$-elements (O and
Mg) and to C, for the lowest and the highest value of
$Z_{\mbox{PopIII}}$. The evolutionary behaviour of N with [Fe/H] will
be discussed in a separate subsection. 
For the lowest values of $Z_{\mbox{PopIII}}$ (which correspond to very
short PopIII phases) the variations of the calculated trends in models
with PopIII relative to our basic model are only observable at the
lowest values of [Fe/H], as we would expect.

If we arise the value of $Z_{\mbox{PopIII}}$, obviously major
variations occur. For a given element, some models can give a better 
result than the CRM03 model: for example, if we consider oxygen,
models H1b and U1Pb seem to fit the observations better. However, the
same models give results at variance with observations for the other
$\alpha$-elements. Moreover, as recently shown by Fran\c cois et
al. (2004) a better fit for oxygen can be obtained  by adopting the
oxygen yields of Woosley and Weaver (1995) as a function of
metallicity. The same model also improves the agreement between
calculations and observations for Mg at low [Fe/H].
On the contrary, PopIII models seem to better represent the Mg data at low
[Fe/H], but fail in reproducing them at higher [Fe/H].

In conclusion, if the primordial PopIII phase was very short, 
as seems likely given the low suggested threshold metallicity, the
results do not allow us to accept or refute models with PopIII and
PCSNe.  If instead it lasted longer, the overall trends are better
reproduced without invoking PopIII. 

\subsection{Models without PCSNe}
\label{sub:noPC}

Here we examine what happens in these PopIII
models which do not involve the formation of PCSN progenitors. These
models have an IMF whose upper limit is given by $M_U=80M_{\odot}$ and
include all the models with yields of CL (from C1 to C4 in Table
\ref{tab-mod1}) and half the models with yields of UN (from U1M to
U4Mb in Table \ref{tab-mod1}).
In Table \ref{tab-res5} are again shown the solar abundances by mass
predicted by the various models compared to the 
outcome of our fiducial model. The last two rows show the duration of
PopIII phase and the number of supernova explosions occurred during
the PopIII phase (where present).
As expected, also in this case, given the short duration of the PopIII
phase, all variations are minimal. Since no change is seen in those
quantities calculated at the present time with respect to our fiducial
model, these results are not presented here.

As to the solar abundances by mass, the same considerations we made in
the previous subsection hold, i.e. we see remarkable differences only
for Fe, Mg and Si. Models with yields of UN underproduce Fe and Si,
and overproduce Mg with respect to the fiducial model, if black hole
formation is not taken into account; when the latter is assumed, all
of the three elements are underproduced relative to the standard model. 
If CL yields are adopted, Mg is overproduced and both Fe and Si are
underproduced. Again, none of these models gives a globally better or
worse agreement with data. We stress that even if the carbon yields of
CL were particularly high, the solar abundance of carbon is dominated
by the production by low and intermediate mass stars which occurs in
the era following PopIII; so, no variation is seen in the results for
carbon. We can notice that  in the models U1Mb$-$U4Mb, since a large
fraction of the forming starsgoes into black holes, the metal
enrichment is particularly slow and, for a given critical metallicity,
the time duration of PopIII phase is about twice the value for other
models; such time duration remains very short anyway.

\begin{table*}
\caption{Solar abundances by mass predicted by models with PopIII
  which do not include PCSNe in their IMF, with yields of Umeda \&
  Nomoto (2002), matched with the results of our fiducial model and
  the observations. Data for solar abundances are from Asplund et
  al. (2004). $\Delta t_{PopIII}$ is the time duration of PopIII phase
  (from the beginning of star formation) and SN\# is  the approximate
  number of supernovae exploded during that phase.}
\label{tab-res5}
\vspace{6pt}
\centering
\begin{center}
\begin{tabular}{lcccccc}
\hline
\hline
$X_i$& Standard& U1M& U2M& U3M& U4M& Obs.   \\
\hline
C& 1.6E$-3$&1.6E$-3$&1.7E$-3$&1.7E$-3$&1.7E$-3$&2.1E$-3$\\
N& 9.6E$-4$&9.6E$-4$&9.6E$-4$&9.6E$-4$&9.6E$-4$&5.9E$-4$\\
O& 5.3E$-3$&5.3E$-3$&5.3E$-3$&5.3E$-3$&5.3E$-3$&5.1E$-3$\\
Mg& 2.5E$-4$&3.8E$-4$&3.8E$-4$&3.8E$-4$&3.8E$-4$&5.7E$-4$\\
Si& 8.3E$-4$&8.2E$-4$&8.1E$-4$&8.1E$-4$&8.1E$-4$&6.3E$-4$\\
Fe& 1.1E$-3$&9.3E$-4$&9.3E$-4$&9.3E$-4$&9.3E$-4$&1.1E$-3$\\
$Z_{\odot}$& 1.2E$-2$& 1.2E$-2$& 1.2E$-2$& 1.2E$-2$& 1.2E$-2$& 1.2E$-2$\\
\hline
$\Delta t_{PopIII}$ ($Myr$)&-&13&4&3&2&-  \\
SN\# & - & 4000 & 150 & 13 & 2 & -  \\
\hline
\hline
\end{tabular}
\end{center}

\centering
\begin{center}

\begin{tabular}{lcccccc}
$X_i$ & Standard & U1Mb& U2Mb& U3Mb& U4Mb& Obs.\\
\hline
C& 1.6E$-3$&1.7E$-3$&1.7E$-3$&1.7E$-3$&1.7E$-3$&2.1E$-3$\\
N& 9.6E$-4$&9.6E$-4$&9.6E$-4$&9.6E$-4$&9.6E$-4$&5.9E$-4$\\
O& 5.3E$-3$&5.4E$-3$&5.3E$-3$&5.3E$-3$&5.3E$-3$&5.1E$-3$\\
Mg& 2.5E$-4$&2.1E$-4$&2.1E$-4$&2.1E$-4$&2.1E$-4$&5.7E$-4$\\
Si& 8.3E$-4$&5.7E$-4$&5.6E$-4$&5.6E$-4$&5.6E$-4$&6.3E$-4$\\
Fe& 1.1E$-3$&9.0E$-4$&9.0E$-4$&9.0E$-4$&9.0E$-4$&1.1E$-3$\\
$Z_{\odot}$& 1.2E$-2$& 1.1E$-2$& 1.1E$-2$& 1.1E$-2$& 1.1E$-2$& 1.2E$-2$\\
\hline
$\Delta t_{PopIII}$ ($Myr$)&-&25&8&6&5&-\\
SN\# & - & 10000 & 400 & 90 & 20 & - \\
\hline
\hline
\end{tabular}
\end{center}
\end{table*}

\begin{table*}
\caption{Same as table \ref{tab-res5} but with yields of Chieffi \&
  Limongi (2002, 2004).}
\label{tab-res7}
\vspace{6pt}
\centering
\begin{center}

\begin{tabular}{lcccccc}
\hline
\hline
$X_i$ & Standard & C1 & C2 & C3 & C4 & Obs.\\
\hline
C& 1.6E$-3$&1.6E$-3$&1.7E$-3$&1.7E$-3$&1.7E$-3$&2.9E$-3$\\
N& 9.6E$-4$&9.6E$-4$&9.6E$-4$&9.6E$-4$&9.6E$-4$&5.9E$-4$\\
O& 5.3E$-3$&5.3E$-3$&5.3E$-3$&5.3E$-3$&5.3E$-3$&5.1E$-3$\\
Mg& 2.5E$-4$&3.5E$-4$&3.5E$-4$&3.5E$-4$&3.5E$-4$&5.7E$-4$\\
Si& 8.3E$-4$&6.1E$-4$&6.1E$-4$&6.1E$-4$&6.1E$-4$&6.3E$-4$\\
Fe& 1.1E$-3$&8.9E$-4$&8.9E$-4$&8.9E$-4$&8.9E$-4$&1.1E$-3$\\
$Z_{\odot}$& 1.2E$-2$& 1.1E$-2$& 1.1E$-2$& 1.1E$-2$& 1.1E$-2$& 1.2E$-2$\\
\hline
$\Delta t_{PopIII}$ ($Myr$)&-&15&4&3&2&-\\
SN\# & - & 4970 & 220 & 13 & 3  &-\\
\hline
\hline
& \\
\end{tabular}
\end{center}
\end{table*}

\begin{figure*}
\centering
\includegraphics[width=.84\textwidth,clip,trim=0 290 0 0]{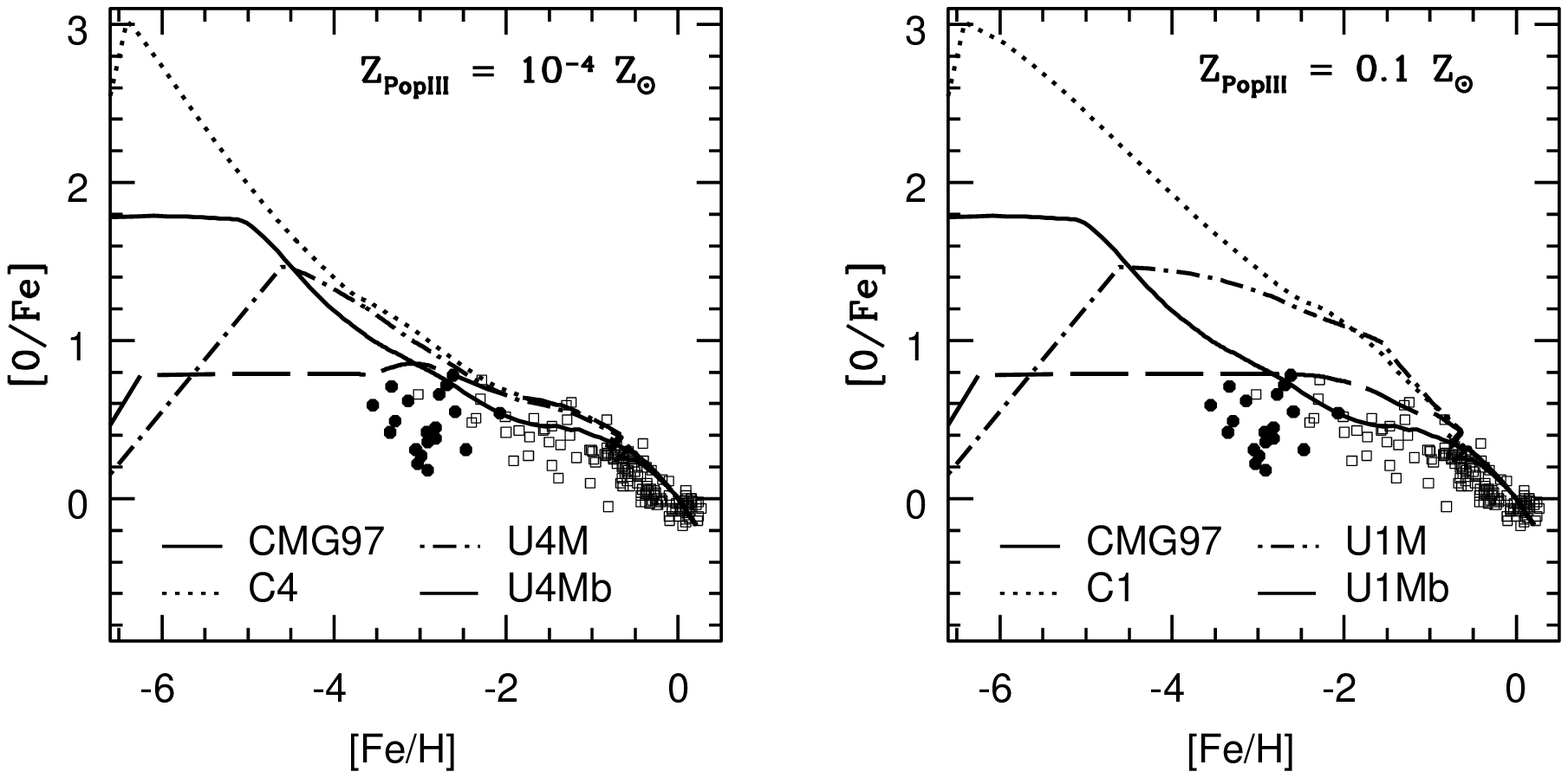}
\includegraphics[width=.84\textwidth,clip,trim=0 290 0 0]{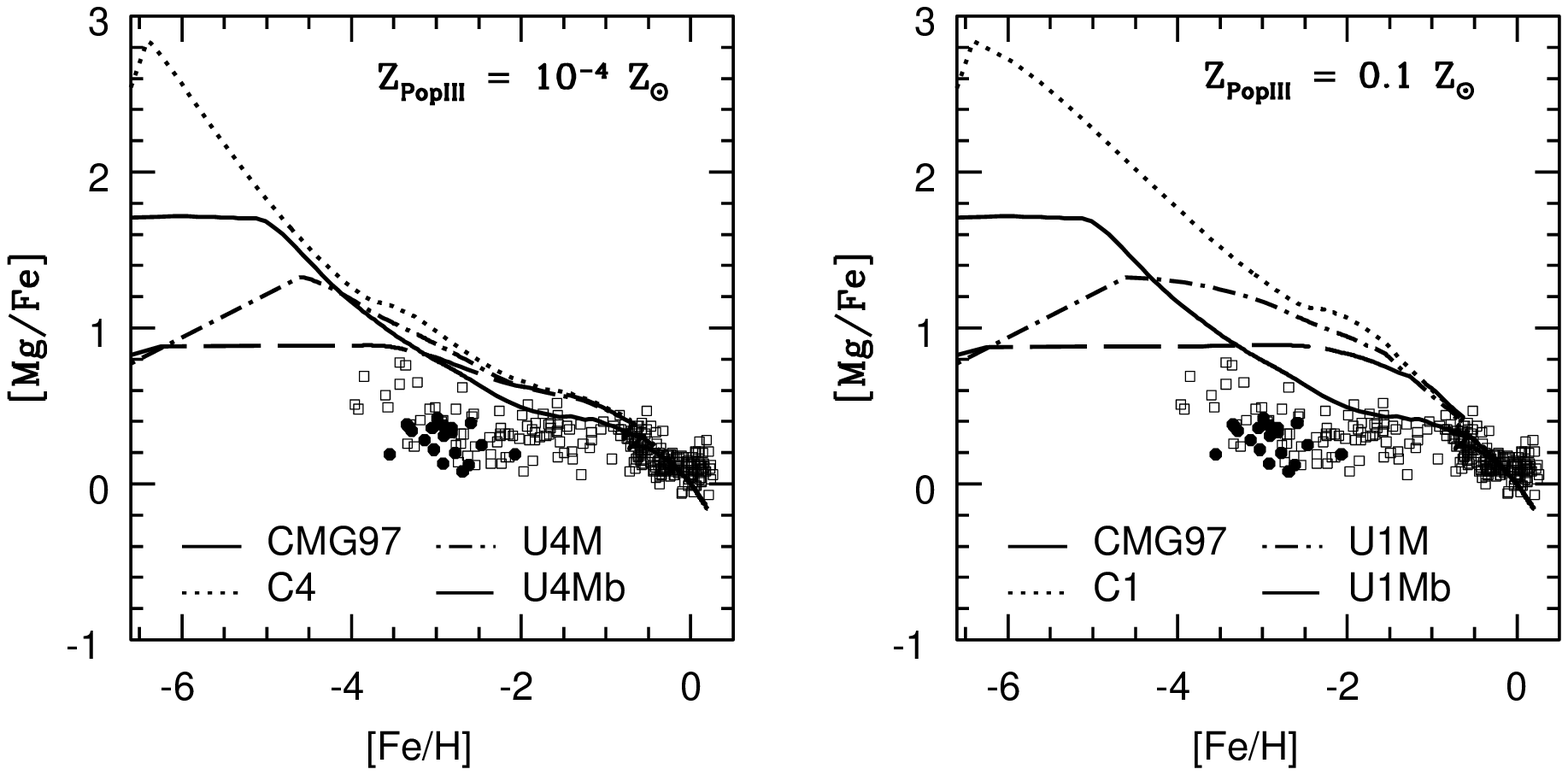}
\includegraphics[width=.84\textwidth,clip,trim=0 290 0 0]{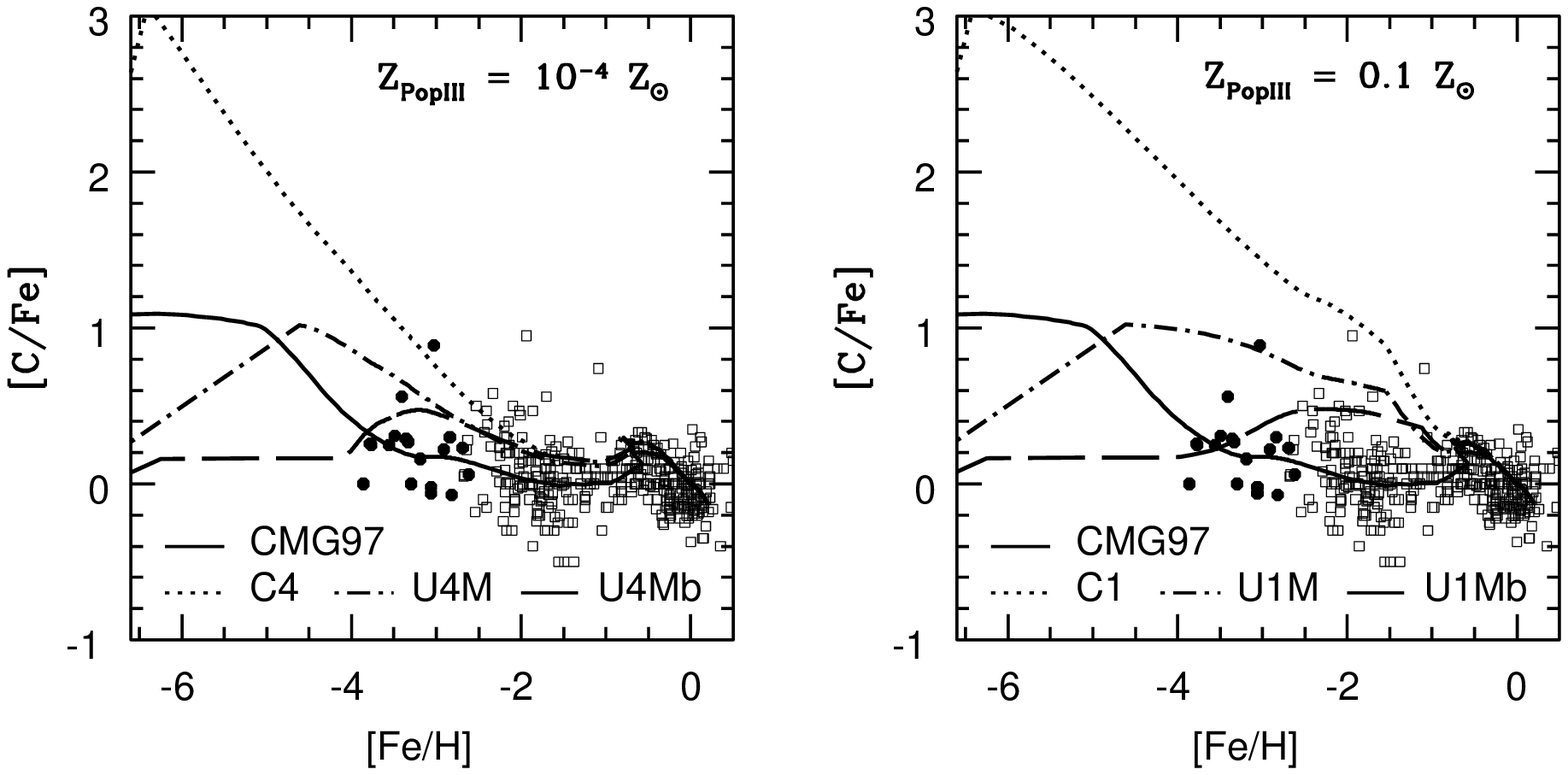}
\caption[]{Evolution of $[$O/Fe$]$, $[$Mg/Fe$]$ and $[$C/Fe$]$ with
  [Fe/H] in our fiducial model (solid line)
  and in the models with PopIII and with IMF truncated 
  at $80M_{\odot}$, thus not including PCSNe included in the IMF, for
  the lowest and highest considered values of the critical
  metallicity. Data are from Chiappini et al. (1999, and references
  therein) for the disk phase (open squares), and for the halo phase
  (filled circles) from Cayrel et al. (2004) for O and Mg and from
  Spite et al. (2005) for C. The plots were extended down to [Fe/H] =
  $-6.5$ in order to show the differences which arise at the lowest
  values of [Fe/H] and which could, in principle, allow one to
  discriminate among the different models once future data are
  available. Please note that the flat behaviour of the  standard
  model of Chiappini et al. (2003) in the  [Fe/H] range between $-4.0$
  and $-6.5$ is due to the extrapolation of the yields of Woosley \&
  Weaver (1995) to masses up to 80 $M_{\odot}$, as described in
    Fran\c cois et al. (2004).} 
\label{fig 3}
\end{figure*}

Fig. \ref{fig 3} shows the evolution with
[Fe/H], for the lowest and the highest values of $Z_{\mbox{PopIII}}$
considered, of the same abundance ratios considered in the previous
subsection, i.e., the trend of $[$X/Fe$]$ vs. $[$Fe/H$]$ for O, Mg, and
C (we shall discuss N in the
following subsection).
Here as well, the longer the PopIII phase lasts, the more
dramatic its effect is, so that none of these models is able to
reproduce the observations for these elements if we adopt $Z_{\mbox{PopIII}}
= 0.1 Z_{\odot}$, with the exception of model U1Mb, which however fails to
reproduce data for Mg in the whole range of [Fe/H]. For the lowest
values of the critical metallicity instead, in most cases only slight
variations are seen with respect to our fiducial model.
A remarkable exception is represented by carbon with the yields of
CL. The calculated trend shows an excessively steep increase of the
$[$C/Fe$]$ ratio towards the lowest values of [Fe/H], which is results
from both the low Fe yields from massive stars and the low
$^{12}C(\alpha,\gamma)^{16}O$ reaction rate assumed by these authors. 

In general, excluding the PCSNe from the IMF does not lead to a better
agreement with observations: the results are still better for a model
which does not include PopIII stars.  
It is worth noting that these models predict a minimum [Fe/H]$ \simeq -6.5$ dex
in the halo gas. On the other hand,  in the models with PCSNe the
predicted minimum [Fe/H] $\sim -4.0$ dex, even higher than the most
metal poor star known (Frebel et al. 2005). In the models of Figure 3
we can see that for [Fe/H]$ < -4.0$ dex remarkable differences among
the models are seen also for those cases with $Z_{\mbox{PopIII}} =
10^{-4} Z_{\odot}$. Forthcoming surveys at very low metallicities may
thus in principle help in putting constraints on the existence of
PopIII in the earliest phases of the Galaxy. However, down to
[Fe/H] $\simeq -4$ the relatively small spread of available data
(e.g. Cayrel et al., 2004; Spite et al., 2005) seems to support our
hypothesis of instantaneous mixing, while for smaller [Fe/H] there are
no available measurements to prove or disprove such an
assumption. Again, low-metallicity surveys could decide whether 
chemical homogeneities were important or not in the early Galaxy. 

\subsection{The nitrogen problem}

\begin{figure*}
\centering
\includegraphics[width=.92\textwidth,clip,trim=0 290 0 0]{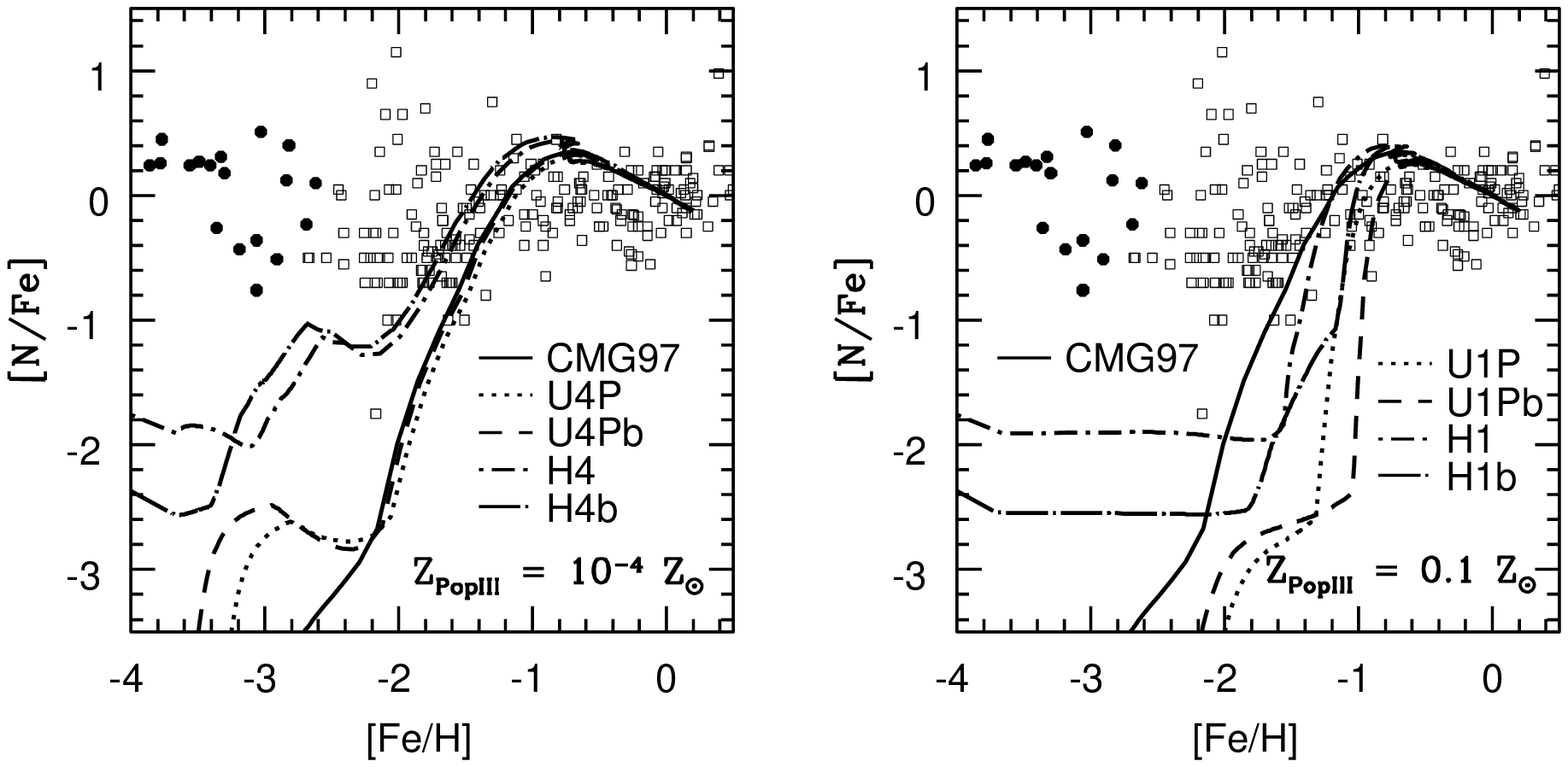}
\includegraphics[width=.92\textwidth,clip,trim=0 290 0 0]{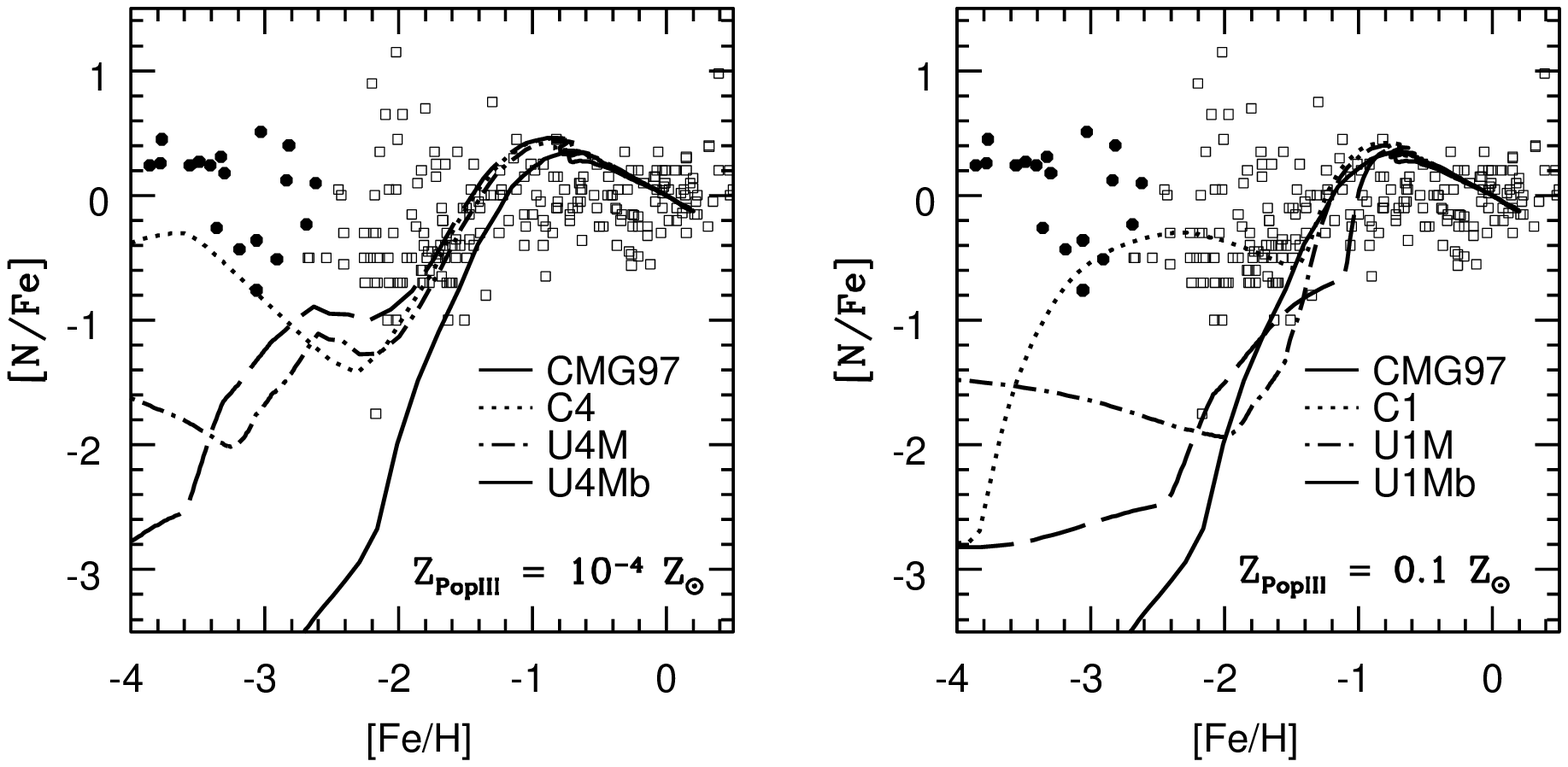}
\caption[]{Evolution of $[$N/Fe$]$ in our fiducial model (solid line)
  and  in the models with PopIII, for the lowest and highest
  considered value of the critical metallicity; here are shown both
  the results of models with PCSNe (upper panels) and without PCSNe
  (lower panels). Data are from
  Chiappini et al. (1999, and references therein) for the disk phase
  (open squares), and from Spite et al. (2005) for the halo phase (filled
  circles).}
\label{fig 4}
\end{figure*}

The main problem with standard models of chemical evolution of the
Galaxy resides in the trend of the relative abundance of nitrogen with
metallicity. These models usually assume that the N production by
massive stars is a secondary process, i.e., N is created starting from
seed nuclei of C already present in the gas out of which the star was
born.
Also low and intermediate mass stars are supposed to produce N in a
secondary fashion but some primary N can be produced in intermediate
mass stars during the dredge-up episodes in conjunction with
hot-bottom burning (Renzini \& Voli, 1981). This implies that the
abundance of nitrogen should increase with metallicity in the earliest
evolutionary phases, and that is what most evolutionary models of the
Galaxy predict. Given the lack of data for N (and the large spread in
those available) until a few years ago, this trend seemed plausible.
However, very recently, high-quality measurements of nitrogen and
carbon abundances appeared (Spite et al., 2005; Israelian et al. 2004). 
These new data indicate that the $[$N/Fe$]$ ratio is constant and
about solar over the whole range of [Fe/H], at variance with the
standard model predictions. One solution to this problem could be
provided by a primary production of N in PopIII stars. We therefore
investigate what is the effect of PopIII on the behaviour
of nitrogen. In Fig. \ref{fig 4} is shown
the evolution of the $[$N/Fe$]$ ratio with [Fe/H] with and
without including PCSN progenitors in the IMF. One can immediately see
that none of the models is able to reproduce the observed constant
trend for all values of [Fe/H], with the exception of model C1 which,
however, yields very unrealistic results for all the other elements
and must therefore be excluded.
In general, all models with PopIII stars without rotation
predict a trend which is somehow approaching the data at the lowest
values of [Fe/H], but if we exclude models with yields of CL, for none
of them we can claim a good agreement. If we adopt a higher value
for $Z_{\mbox{PopIII}}$, in most cases the trend flattens well below
the data. This fact shows that PopIII stars alone cannot provide a
sufficient primary production of N. For some models (e.g. U1P and
U1Pb) the plot of $[$N/Fe$]$ vs. $[$Fe/H$]$ decreases even more
steeply than the CRM03 model towards low [Fe/H]. 

\begin{figure*}
\centering
\includegraphics[width=.83\textwidth,clip,trim=0 290 0 0]{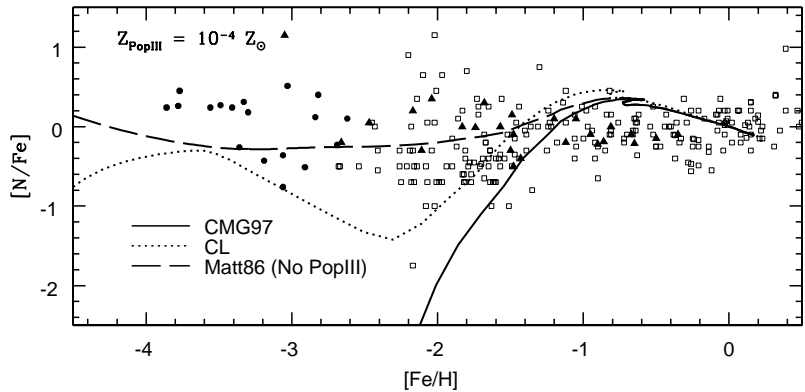}
\caption[]{Evolution with [Fe/H] of the $[$N/Fe$]$ ratio in our
  fiducial model (solid line), in the PopIII model with yields of CL
  and $Z_{\mbox{PopIII}}=10^{-4}Z_{\odot}$ (dotted
  line) and in model Matt86 (dashed line) where a constant and primary
  production of $^{14}$N is imposed at all metallicities for stars of
  all masses. Data are from Chiappini et al. (1999, open squares),
  Spite et al. (2005, filled circles) and Israelian et al. (2004,
  filled triangles).}
\label{fig 5}
\end{figure*}

We now examine more closely the models with yields of CL. The
better agreement shown by these models for the $[$N/Fe$]$ ratio is due
to the behaviour of the N/Fe production ratio of PopIII stars. 
Stars of $30M_{\odot}$ contibute to the chemical enrichment
  when [Fe/H] $\simeq -4$, while  stars of $50M_{\odot}$ explode
  almost immediately after the beginning of star formation, when
  [Fe/H] $\simeq -4.5$. With the yields of CL, the N/Fe production
  ratio for stars of $30M_{\odot}$ is about a factor of 1000 greater
  than that for stars of $50M_{\odot}$  which leads to an increase of
  $[$N/Fe$]$ in the very first Myrs (i.e. between [Fe/H] $= -4.5$ and
  $-4$). \footnote{Of course, this results from the assumption of
    instantaneous mixing, i.e. that the timescale for N to mix with the
    environment is much shorter than the difference between the
    lifetimes of a $30M_{\odot}$ and a $50M_{\odot}$ star. If this
    timescale were longer, then such a behaviour would not appear.}
In Fig. \ref{fig 5} it is clear that PopIII stars alone do not
solve the problem because, apart from the reasonably good value of
$[$N/Fe$]$ at very low [Fe/H], the predicted $[$N/Fe$]$ ratio
decreases afterwards to increase again for [Fe/H] $> -2.0$, as
expected from the normal secondary N production in massive stars.
It is worth noting that even for the lowest values of
$Z_{\mbox{PopIII}}$ these models with CL yields give results for
$[$C/Fe$]$ in contrast with observations.
Finally,  we compare (always Fig. \ref{fig 5}) the results of model C4
with a heuristic model calculated by Matteucci (1986, Matt86); the
latter model is still based on the CRM03 model, but assumes that all
massive stars produce primary nitrogen. In particular, they yield a
constant ($0.065M_{\odot}$) quantity of $^{14}$N at every
metallicity. Although this assumption was made \emph{ad hoc}, this
model seems to fit the data better than the standard model and is
useful to understand that the primary production of nitrogen must
occur at every  metallicity and not only by means of a stellar
population confined to the halo phase such as PopIII (in fact, model
C4 does not reproduce the data at intermediate [Fe/H]). Therefore,
mechanisms other than PopIII must be invoked. So far, we took
into account only PopIII yields computed without rotation. For
example, Meynet \& Maeder (2002) calculated that  stellar rotation as
a function of metallicity can increase the primary production of
nitrogen, and even though it has been shown (Chiappini et al., 2003b)
that their rotation yields are not sufficient to obtain the requested
trend, this is a mechanism that should be further studied. 
In other words, we cannot predict the exact amount of primary nitrogen 
produced by massive stars nor its dependence upon the stellar mass but 
only suggest that a continuous N production from massive stars is
required at every metallicity (see Chiappini et al. 2005, for a more
extensive discussion on the N production).

\subsection{The $[$C/O$]$ ratio}

\begin{figure*}
\centering
\includegraphics[width=.83\textwidth,clip,trim=0 290 0 0]{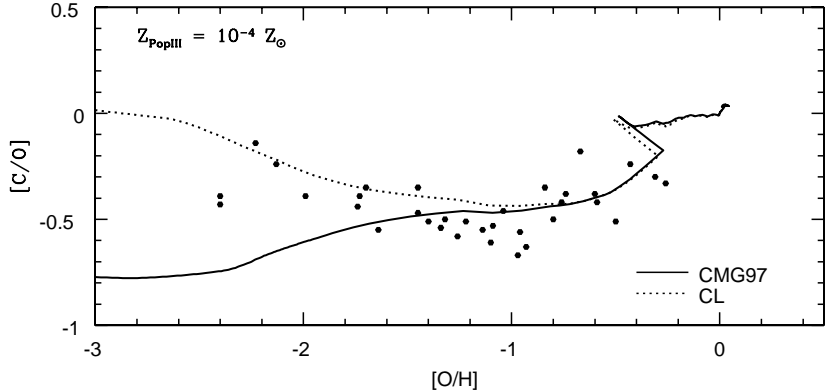}
\caption[]{Evolution of the $[$C/O$]$ ratio with $[$O/H$]$ in the
  fiducial model (solid   line) and in the PopIII model with yields of
  CL and   $Z_{\mbox{PopIII}}=10^{-4}Z_{\odot}$ (dotted line). The data
  are from Akerman et al. (2004)} 
\label{fig 6}
\end{figure*}

Another result for which PopIII has lately been invoked (and, more in
detail, PopIII with yields of CL) is the observed trend of $[$C/O$]$
vs. $[$O/H$]$, which appears to rise towards low [O/H]. In
fact, Akerman et al. (2004) have shown that a chemical evolution model
with the assumption of a PopIII phase lasting until a
metallicity of $10^{-4}Z_{\odot}$ was reached, with an IMF including
only massive stars ($10-80M_{\odot}$) and where CL yields were
adopted (equivalent to our model C4), is able to
reproduce the observed trend, whereas our fiducial model does not
follow the data at low [O/H] (see Fig. \ref{fig 6}). This is mainly
due, in CL yields, to the adopted rate for the reaction
$^{12}C(\alpha,\gamma)^{16}O$ at the end of He-burning in the core of
metal-free stars; the rate being low, the C/O production ratio at low
metallicities is increased.
However, the same authors (Akerman at al., 2004) in their analysis
did not take into account non-LTE corrections which depend on
metallicity and could tend to systematically lower the measured
$[$C/O$]$ ratio. They also did not exclude the possibility that the
apparent trend is an artifact due to limited statistics. Moreover, we
have seen in \S \ref{sub:noPC} that models with yields of CL cannot
reproduce the observations of $[$C/Fe$]$ vs. $[$Fe/H$]$ for the same
reason thanks to which they can well fit the data for $[$C/O$]$
vs. $[$O/H$]$ (i.e., the rate of $^{12}C(\alpha,\gamma)^{16}O$).
Instead, if one
adopts O yields from Woosley \& Weaver (1995) as a function of
metallicity (see Chiappini et al. 2005 and Fran\c cois et al., 2004)
it is possible to increase the $[$C/O$]$ ratio at low [O/H]. Thus,
even if these data were correct, there is not a unique way to
reproduce them, and the models which give the best overall results are
those which do not invoke PopIII.

\section{The effects of a variable IMF}

In this section we briefly explore the effect of a different kind of
IMF which could in principle provide a pre-enrichment of the Galactic
gas. For this purpose, we adopted the IMF suggested by Larson (1998)
which has the following form:

\begin{equation}
\frac{d\varphi(m)}{d\log{m}} \propto \left(1+\frac{m}{M_C}\right)^{-x}
\label{eq:larson}
\end{equation}
where $M_C$ is a ``characteristic'' mass at which the IMF is peaked,
whose scale is given by the Jeans mass of collapsing clouds.
By adequately changing the value of $M_C$ as a function of time, one
can obtain an IMF which is ``top-heavy'' in the earliest phases of the
Galaxy. In this way we built a ``Flat'' model and two ``Hernandez \&
Ferrara'' models. In the ``Flat'' (hereafter F) model, the
peak of the IMF is shifted towards larger masses during the whole halo
phase; this is a heuristic model by means of which we investigate the
behaviour of chemical evolution with a strongly variable
IMF. Referring to Eq. \ref{eq:larson}, for the F model the IMF 
exponent is $x=1.8$ and $M_C$ is assumed equal to $10M_{\odot}$ for
the first 1Gyr (the duration of the halo phase), then it switches to
$0.35M_{\odot}$ and remains constant since then. The choice of $x=1.8$
and $M_C = 0.35M_{\odot}$ for the present-time IMF allows us to
reproduce the observed solar abundances by mass. In the ``Hernandez \&
Ferrara'' (hereafter hfI and hfII) models, the variation of the peak mass
follows the  suggestion made by Hernandez \& Ferrara  (2001); these
authors explore the predictions of the  hierarchical 
model of galaxy formation about the number and metallicities of
metal-poor stars and, comparing these predictions with observational
data and  assuming a Larson IMF, they conclude that the characteristic mass
$M_C$ should increase with redshift.
For the hfI and hfII models, we referred to Fig. 6 of Hernandez \& Ferrara
(2001), where is shown the variation of $M_C$ as a function of
metallicity, from which we extrapolated the following trends for the
characteristic mass:
$$
M_C=
\left\{
\begin{array}{rcl}
10 M_{\odot} &\mbox{\  if \ }&Z/Z_{\odot} \leq 10^{-3.92}\\
13 M_{\odot} &\mbox{\  if \ }&10^{-3.92} < Z/Z_{\odot} \leq 10^{-3.77}\\
14 M_{\odot} &\mbox{\  if \ }&10^{-3.77} < Z/Z_{\odot} \leq 10^{-3.58}\\
11 M_{\odot} &\mbox{\  if \ }&10^{-3.58} < Z/Z_{\odot} \leq 10^{-3.38}\\
6.5 M_{\odot}&\mbox{\  if \ }&10^{-3.38} < Z/Z_{\odot} \leq 10^{-2.9}\\
2.5 M_{\odot}&\mbox{\  if \ }&10^{-2.9}  < Z/Z_{\odot} \leq 10^{-2.55}\\
0.8 M_{\odot}&\mbox{\  if \ }&10^{-2.55} < Z/Z_{\odot} \leq 10^{-2.2}\\
0.35 M_{\odot}&\mbox{\  if \ }&Z/Z_{\odot} > 10^{-2.2}
\mbox{ in Model I} \\
0.50 M_{\odot}&\mbox{\ if \ }&Z/Z_{\odot} > 10^{-2.2}
\mbox{ in Model II}
\end{array}
\right.
$$
Moreover, $x=1.35$ for Model I and $x=1.8$ for Model II. These choices
are due to the fact that Model I exactly reproduces the IMF suggested
in the paper, whereas in Model II, while still adopting the same variation
as a function of metallicity, the parameters $x$ and $M_C$ are chosen
in order to reproduce the observed solar abundances by mass.

The basic model is still the CRM03 best model; excluding the IMF,
all the other prescriptions were left unchanged. In particular, the
same nucleosynthetic prescriptions hold as in \S \ref{sec:mod} (i.e.,
the yields adopted for massive stars in this case are those of Woosley
\& Weaver (1995), solar composition, at all times).

\begin{table*}
\centering
\caption{Predictions for models with Larson IMF in the solar neighbourhood
matched with the results of the CRM03 model and the
observations. Among the quantities calculated at the present time
(in brackets are indicated the data sources): $\Psi_0$ is the  star
formation rate (Rana, 1991); $\mathcal{F}$$_0$ is the infall rate
(Portinari et al., 1998); $\sigma_g$ is the gas surface mass density
(Dickey, 1993); $\sigma_{\star}$ is the star surface mass density
(Gilmore et al., 1989); $\sigma_{\star,rem}$ is the surface mass
density of stars plus remnants (M\'era et al., 1998); $\sigma_{tot}$
is the total (gas + stars + remnants) surface mass density (Sackett,
1997); finally, $R_{SNII}$ is the present time rate of Type II SN explosions
(Tammann et al., 1994).}
\label{tab:varmod}
\vspace{6pt}
\begin{tabular}{l r@{.}l r@{.}l r@{.}l r@{.}l c}
\hline 
\hline
Quantity  & 
\multicolumn{2}{c}{F}  & 
\multicolumn{2}{c}{hfI}  & 
\multicolumn{2}{c}{hfII} & 
\multicolumn{2}{c}{Standard} & Obs.\\
\hline
$\Psi_0$ ($M_{\odot}pc^{-2}Gyr^{-1}$) & 3&1  & 3&8  & 3&1  & 2&6 
& $2-5$\\
$\mathcal{F}_0$ ($M_{\odot}pc^{-2}Gyr^{-1}$) & 1&1 & 1&1 & 1&1 & 1&02
& $0.3-1.5$\\
$\sigma_{g}$ ($M_{\odot}pc^{-2}$) & 7&0  & 8&1  & 7&0  & 7&1
& $7.0$ \\
$\sigma_{\star}$ ($M_{\odot}pc^{-2}$) & 43&3 & 35&6 & 38&5 & 35&4  
& $35\pm5$\\
$\sigma_{\star,rem}$ ($M_{\odot}pc^{-2}$)& 44&8 & 43&7 & 44&8& 42&9 
&$43 \pm 5$\\
$\sigma_{tot}$ ($M_{\odot}pc^{-2}$) & 51&8 &  51&8 & 51&8 & 50&0 
&$51\pm6$\\
$R_{SNII}$ ($pc^{-2}Gyr^{-1}$) & 0&018 & 0&056 & 0&019 & 0&014 
&$0.020$\\
\hline
\hline
\end{tabular}
\vspace{9pt}
\caption{Solar abundances by mass calculated by models with a Larson-type
  variable IMF  matched with the results of our fiducial model and the 
  observations.  Data for solar abundances are from Asplund et
  al. (2004).} 
\label{tab:abb}
\vspace{6pt}
\begin{tabular}{lcccccc}
\hline 
\hline
$X_i$ & F  & hfI  & hfII & Standard & Obs.\\
\hline
C         &$2.6$E$-3$ &$5.1$E$-3$ & $2.9$E$-3$& $1.6$E$-3$& $2.1$E$-3$\\
N         &$2.2$E$-3$ &$6.2$E$-3$ & $2.1$E$-3$& $9.6$E$-4$& $5.9$E$-4$\\
O         &$1.0$E$-2$ &$2.9$E$-2$ & $1.1$E$-2$& $5.3$E$-3$& $5.1$E$-3$\\
Mg        &$4.2$E$-4$ &$1.4$E$-3$ & $4.1$E$-4$& $2.5$E$-4$& $5.7$E$-4$\\
Si        &$1.0$E$-3$ &$2.7$E$-3$ & $1.1$E$-3$& $8.3$E$-4$& $6.3$E$-4$\\
Fe        &$1.4$E$-3$ &$3.0$E$-3$ & $1.6$E$-3$& $1.1$E$-3$& $1.1$E$-3$\\
$Z_{\odot}$&$2.1$E$-2$&$5.9$E$-2$ & $2.1$E$-2$& $1.2$E$-2$& $1.9$E$-2$\\
\hline
\hline
& \\
\end{tabular}
\end{table*}

Table \ref{tab:varmod} shows the results concerning various quantities
calculated at the present time (i.e., the Galactic age $t_G=13.7$ Myr)
for all the models presented in this section, compared to the results
of the basic model and to the observations. With a few exceptions, model
results generally agree with the data. Small discrepancies are seen in
model F which tends to overestimate the surface mass density of stars,
while model hfI strongly overestimates the present-day rate of
supernova explosions. This is due to the choice of $x=1.35$ which
results in a too flat IMF if compared with other IMFs (Scalo, 1986;
Chabrier, 2003) which better reproduce the features of the solar
neighbourhood. Such a flat IMF leads to an overproduction of supernova
progenitors even at the present time. 

In Table \ref{tab:abb} we show the calculated solar abundances by
mass. If we exclude model hfI, all the predicted values are within a
factor of 2 of the observed ones and thus acceptable (with the
exception of N). The strong overproduction of metals in model hfI can 
again be explained by the excessive flatness of the chosen IMF, which
favours the overproduction of massive stars.

\begin{figure*}
\centering
\includegraphics[width=.33\textwidth, clip, trim=5 290 290 0]{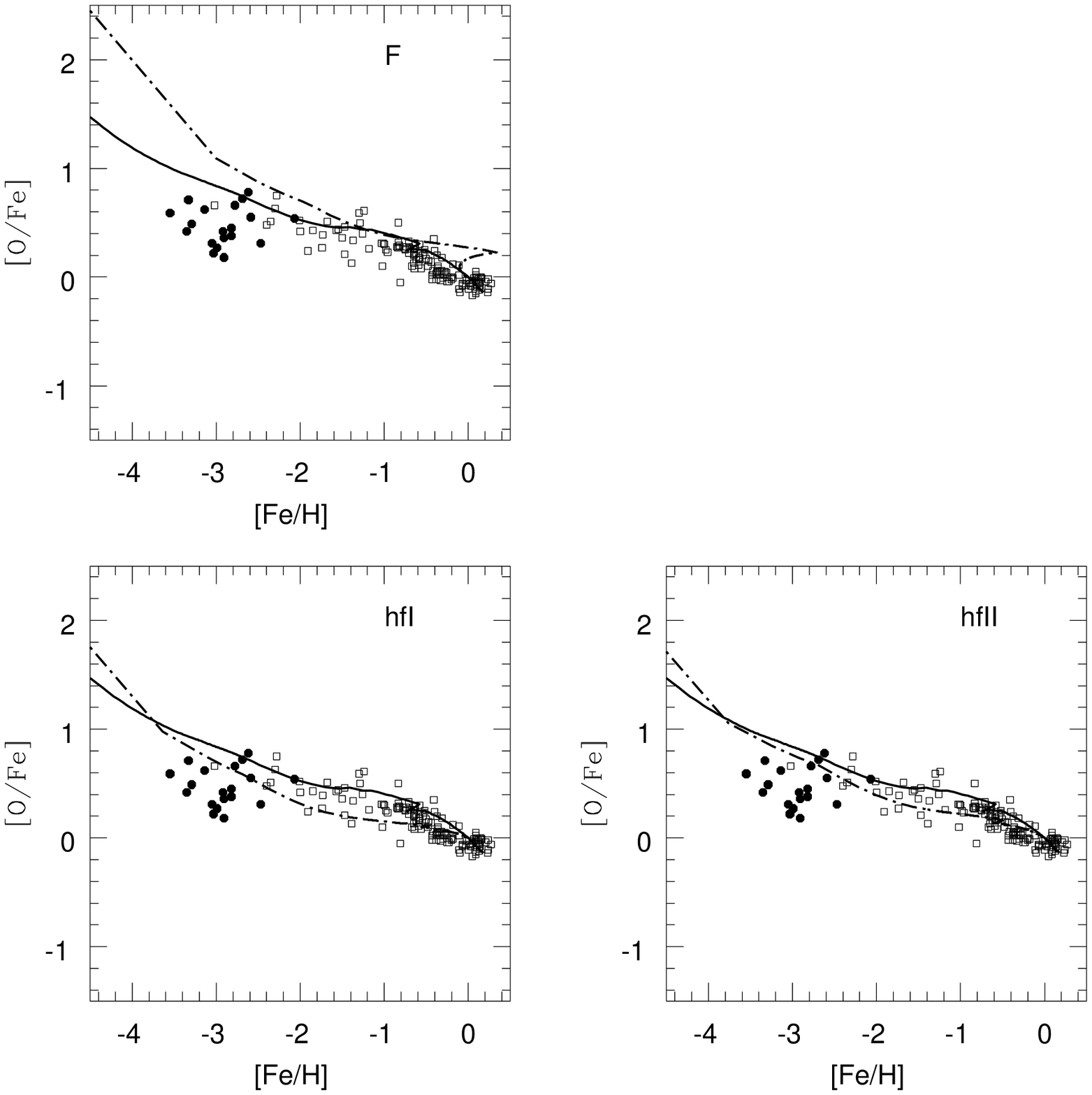}%
\includegraphics[width=.66\textwidth, clip, trim=5 18 0 290]{Fig7A.eps}
\includegraphics[width=.33\textwidth, clip, trim=5 290 290 0]{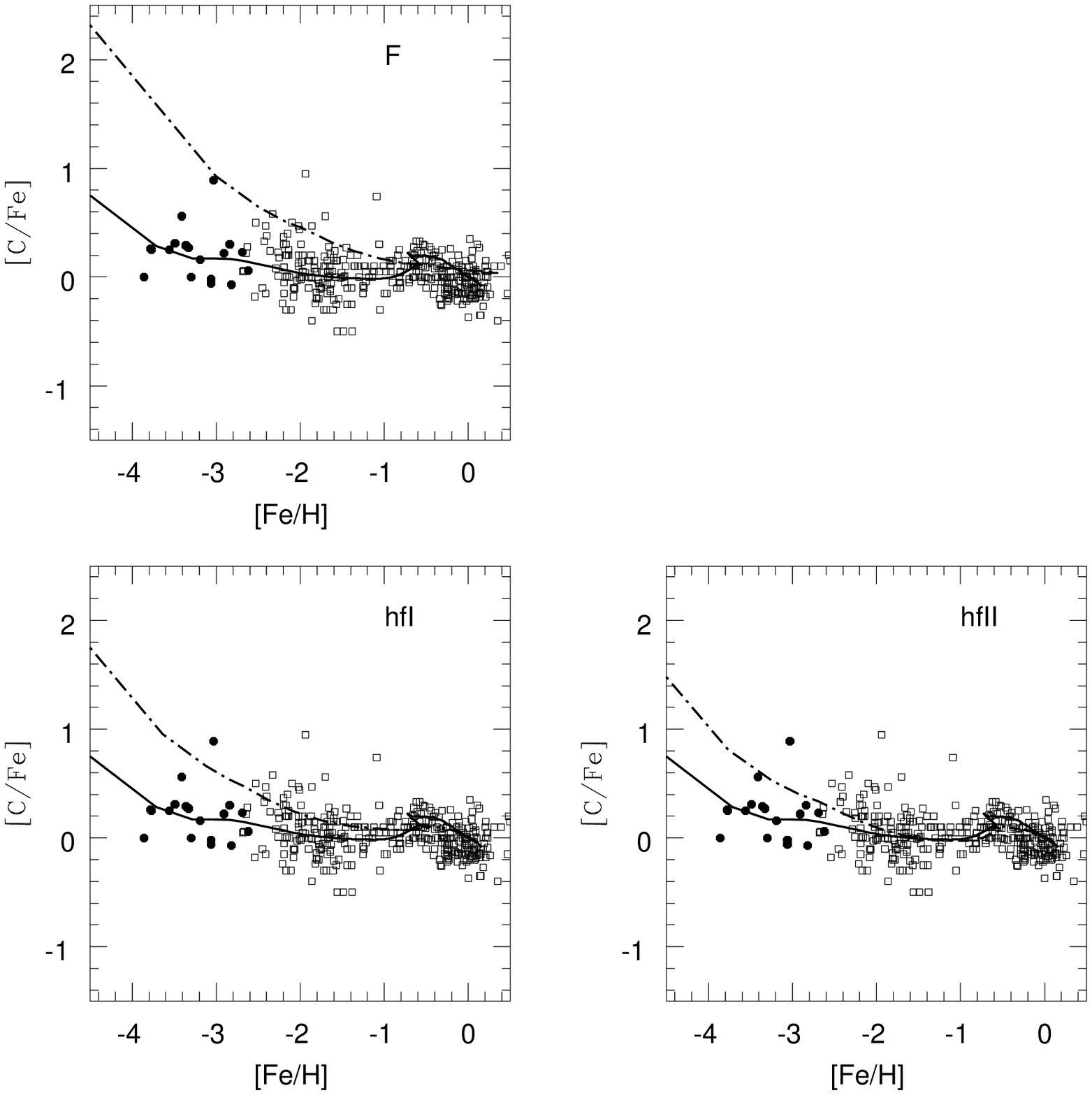}%
\includegraphics[width=.66\textwidth, clip, trim=5 18 0 290]{Fig7B.eps}
\includegraphics[width=.33\textwidth, clip, trim=5 290 290 0]{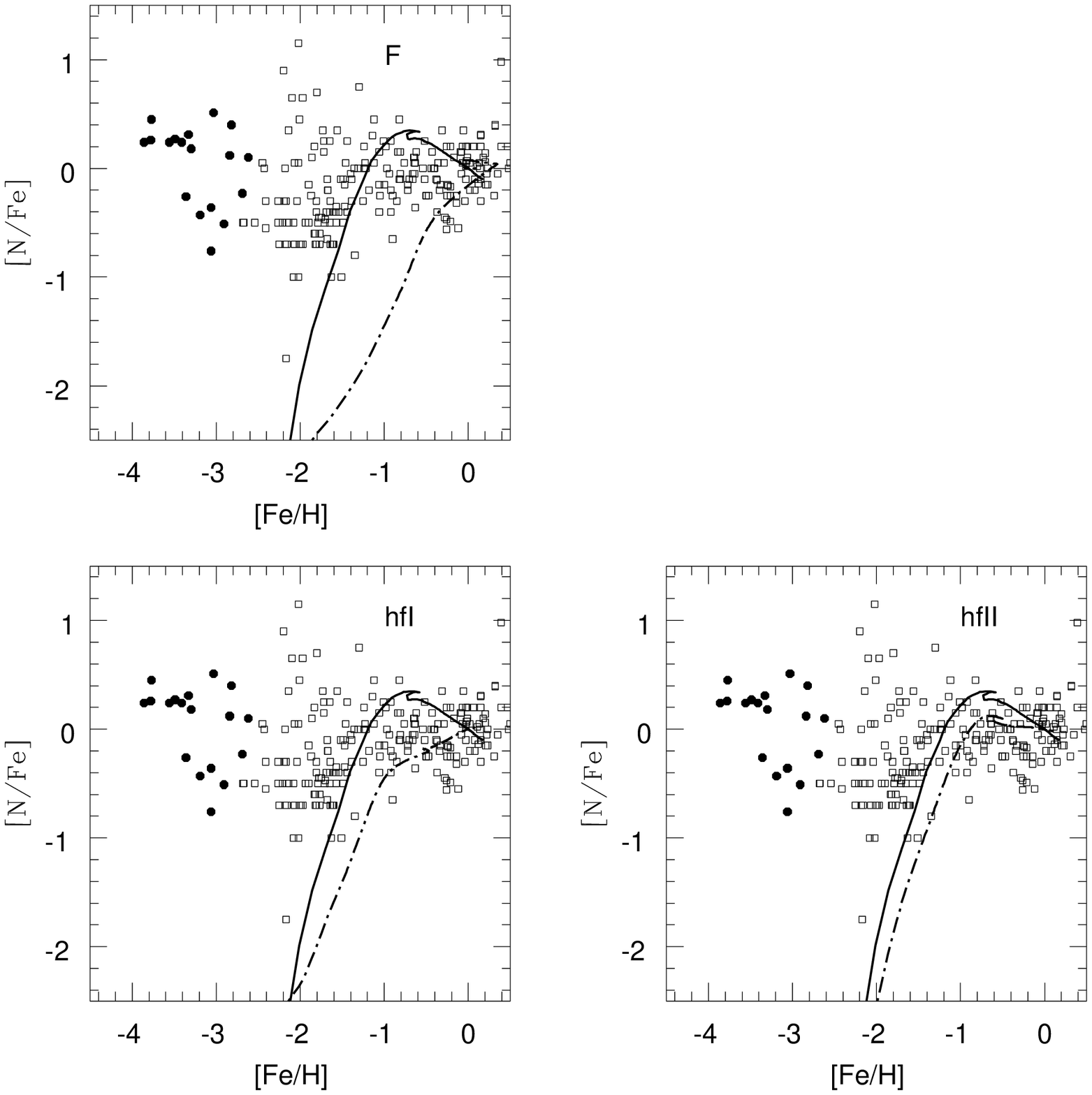}%
\includegraphics[width=.66\textwidth, clip, trim=5 18 0 290]{Fig7C.eps}
\caption{Evolution of $[$O/Fe$]$, $[$C/Fe$]$ and $[$N/Fe$]$  with
  [Fe/H] in the ``Halo'' models with variable IMF (dashed-dotted line)
  compared to the results predicted by model CRM03 (solid line). Data
  for the disk phase (open squares) are from Chiappini et al. (1999)
  and references therein; data for the halo phase (filled circles) are
  from Cayrel et al. (2004) for O, from Spite et al. (2005) for C and N.} 
\label{fig 7}
\end{figure*}

Fig. \ref{fig 7} shows the evolution of the
abundance ratios $[$O/Fe$]$, $[$C/Fe$]$ and $[$N/Fe$]$ with
[Fe/H] predicted by the models considered here. Model F severely
overproduces oxygen and carbon and underproduces nitrogen even at
intermediate [Fe/H]. Model hfI predicts an excessively steep
trend of the carbon abundance towards low [Fe/H] and slightly
underproduces nitrogen. Similar trends are predicted by model hfII
which however presents smaller differences relative to the basic
model and thus cannot be ruled out on the basis of abundance analysis,
although the best overall agreement is still obtained by a constant
IMF. The behaviour of other $\alpha$-elements (Mg, Si) is very similar
to  that of oxygen, so the same conclusions hold.

\begin{figure*}
\centering
\includegraphics[width=.44\textwidth,clip,trim=0 290 290 0]{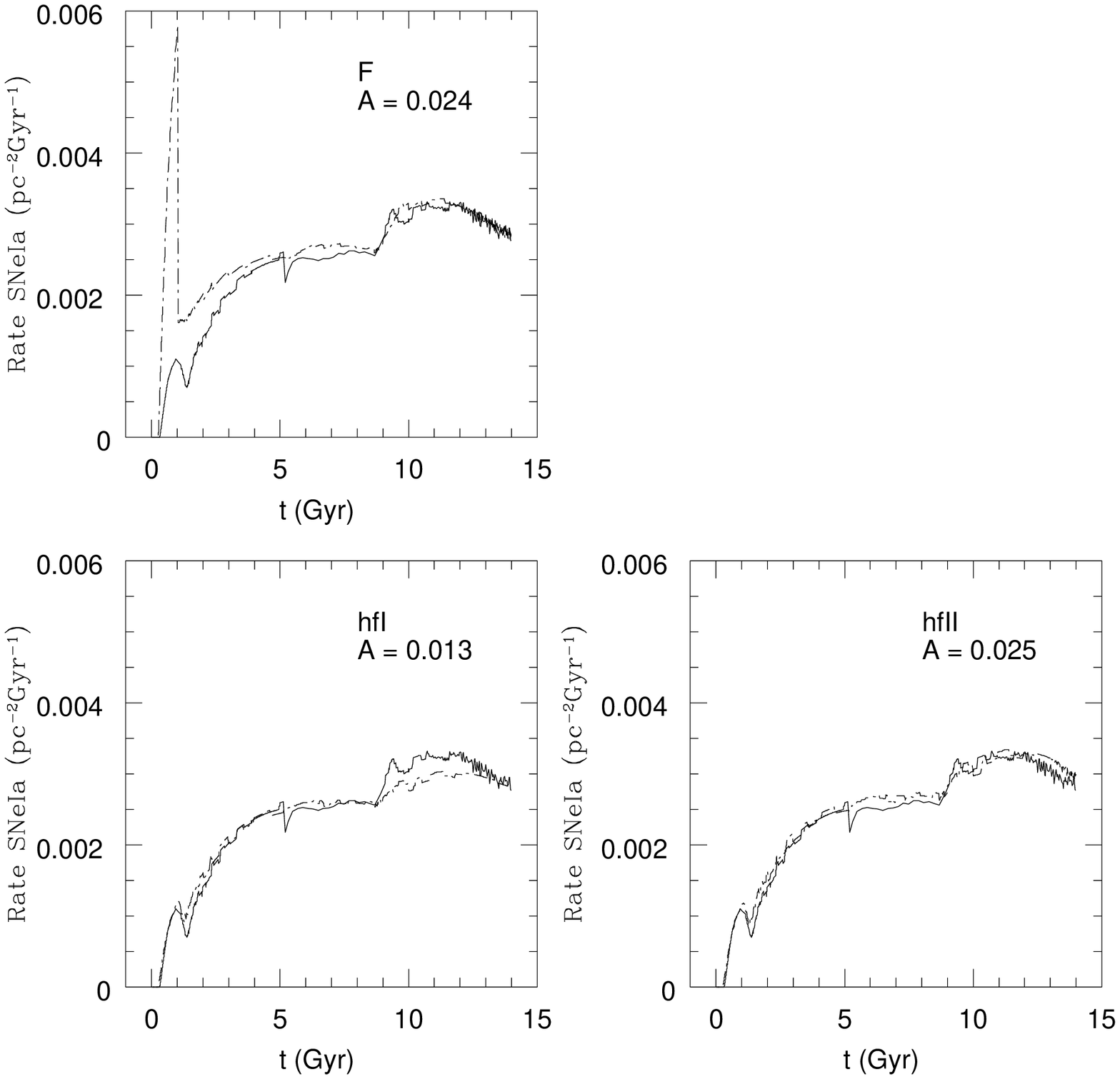}
\caption{Evolution of Type Ia SN rate  in the ``Halo'' models with
  variable IMF (dashed-dotted line) compared to the results predicted
  by model CRM03 (solid line). $A$ is the assumed fraction of binary
  systems in the IMF which evolve to Type Ia SNe.}
\label{fig 8}
%
\centering
\includegraphics[width=.33\textwidth, clip, trim=0 290 291 0]{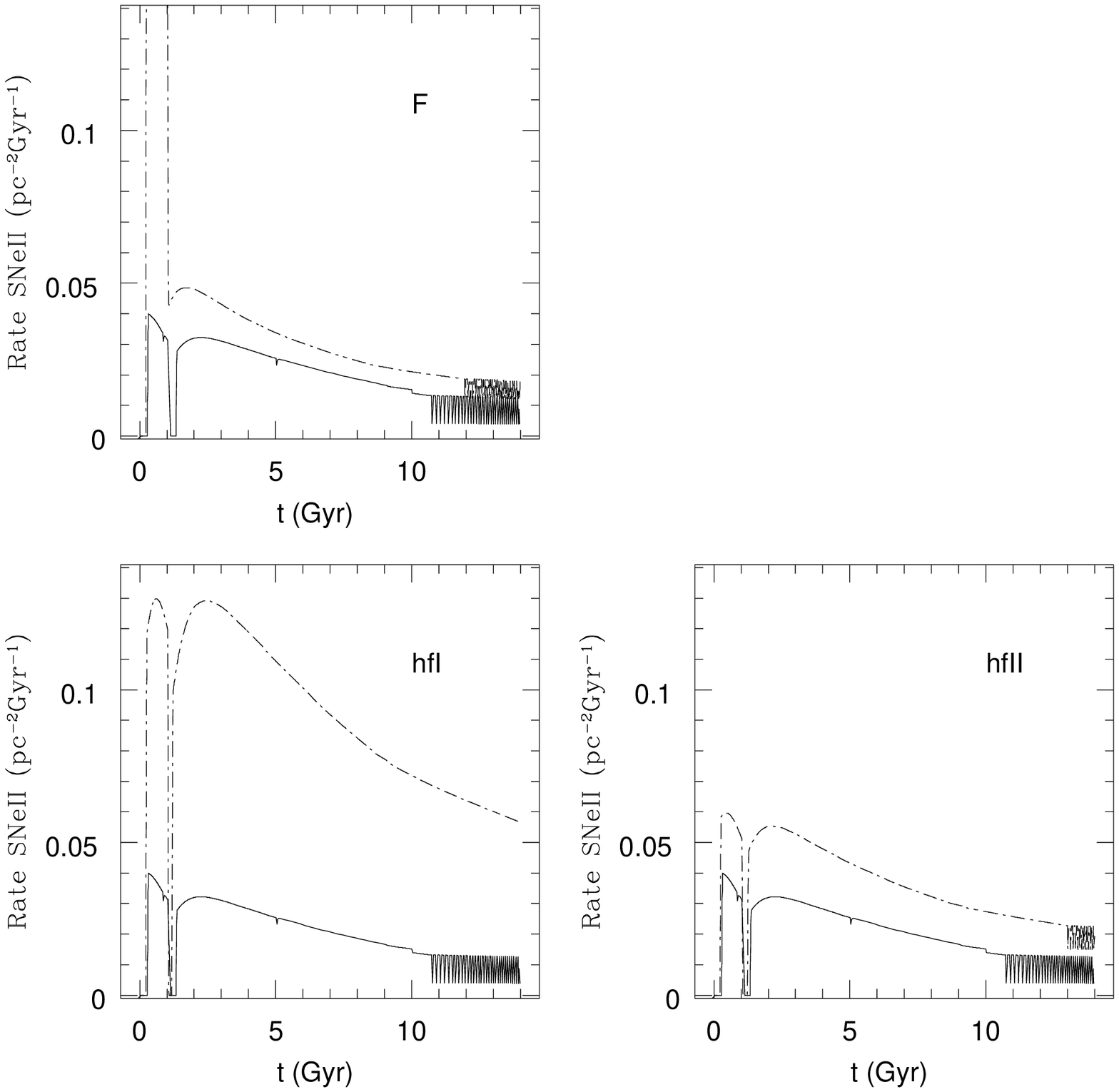}%
\includegraphics[width=.66\textwidth, clip, trim=0 18 1 290]{Fig9.eps}
\caption{Evolution of Type II SN rate for all the models with variable
  IMF (dashed-dotted line) compared to the results predicted
  by model CRM03 (solid line). The peak in the halo phase was cut off
  in model F for graphical reasons. The dark regions seen after 10 Gyr
  are due to the fast oscillations of SFR when the threshold surface
  mass density is reached (see Chiappini et al., 1997).}
\label{fig 9}

\includegraphics[width=.33\textwidth, clip, trim=0 290 290 0]{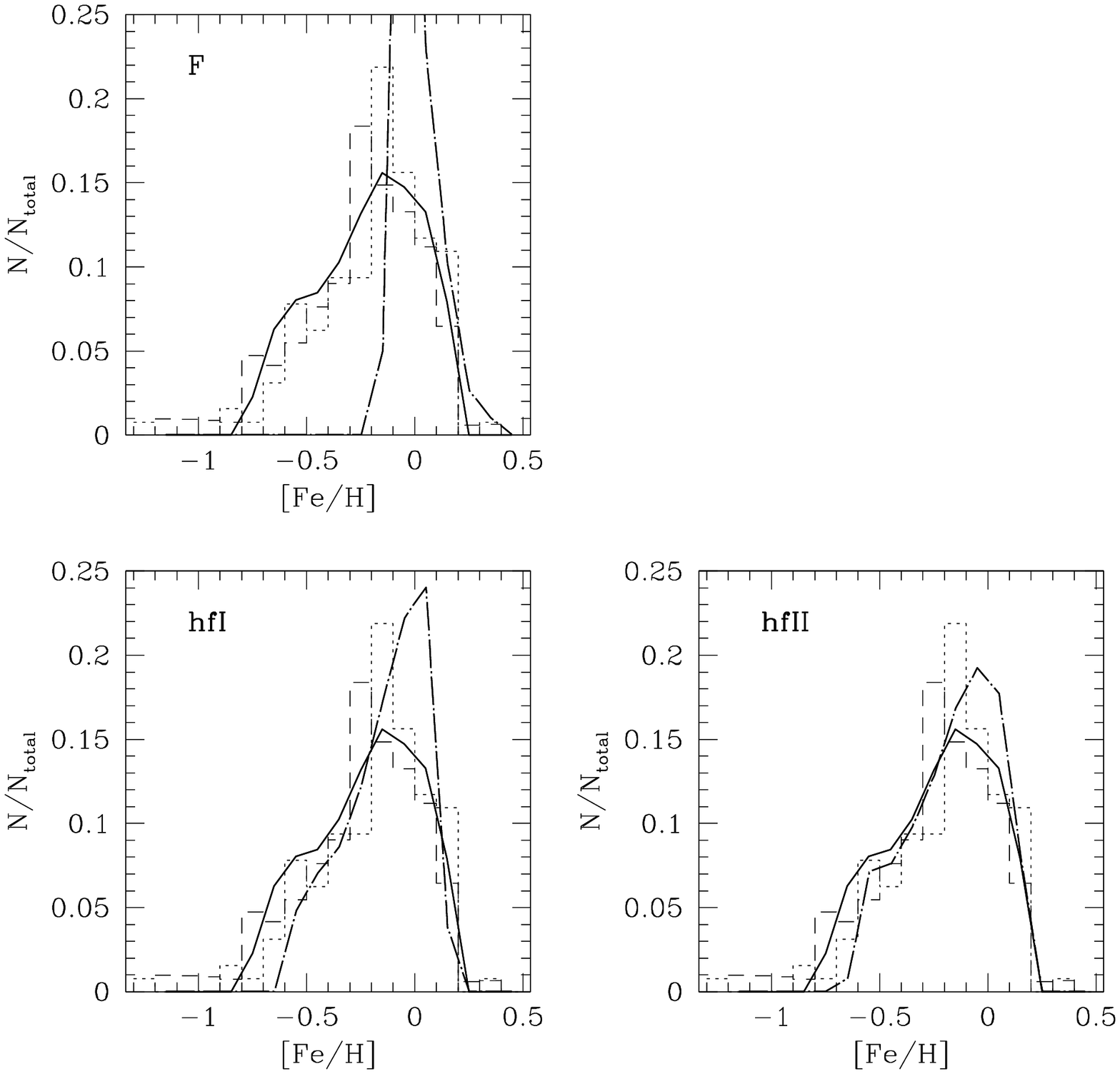}%
\includegraphics[width=.66\textwidth, clip, trim=0 12 0 290]{Fig10.eps}
\caption{G-dwarf metallicity distribution in the solar neighbourhood
  for our fiducial model (solid line) and for the models with variable
  IMF (dashed-dotted line). Data are from Wyse \& Gilmore (1995, dashed
  hystogram) and Rocha-Pinto \& Maciel (1996, dotted hystogram).}
\label{fig 10}
\end{figure*}

Fig. \ref{fig 8} shows the temporal evolution of the SNIa rate for model
F (the other models do not show remarkable differences relative to
the basic model). We can see the pronounced peak during the halo phase
which indicates that, because we chose $M_C=10M_{\odot}$ as the ``most
probable'' mass in the IMF for the halo phase, the number of Type Ia SN
progenitors, which are stars of intermediate mass, is increased by the
adoption of this kind of IMF. Such a result also demonstrates how
critically the results may depend on the details of the adopted IMF:
in fact, the IMF being peaked near the intermediate mass range, the N
production is enhanced for all models. 
The sharp feature in SNIa rate is not seen in models hfI and hfII,
owing to the extremely limited time duration of the ``top-heavy''
phase which ends after 230~Myr.

In Fig. \ref{fig 9} is shown the temporal evolution of the SNII rate
for all the models presented here. We notice that the present-day SNII 
rate crucially depends on the IMF exponent, as already mentioned; this
can be seen by the differences between model hfI and the other models
which have $x=1.8$.

Finally, Fig. \ref{fig 10} shows the metallicity distribution of
G-dwarfs in the solar neighbourhood predicted by these models; model F
and model hfI are not able to fit the data well, predicting an
extremely sharp distribution peaked at too high [Fe/H]. On the
other hand, model hfII shows an acceptable agreement with data;
however, the distribution peak is still located at [Fe/H]
higher than that observed. 

In conclusion, the effects of a variable IMF which is ``top-heavy'' in
the earliest phases result in an overproduction of metals, an increasing
slope for the evolution of abundance ratios towards low [Fe/H]
and a shift towards larger [Fe/H] of the G-dwarf distribution
peak. Depending on the particular chosen IMF, there can also be
different results in abundance evolution which largely depend on the
adopted nucleosynthetic prescriptions. Most results of a variable IMF
are at variance with observations, and an overall better agreement is
achieved with a constant IMF, unless the suggested IMF variations are
so small that they are irrelevant (see also Chiappini et al., 2000).

\section{Summary and conclusions}
In this paper we have explored the effects on the chemical evolution
of the solar neighbourhood of one or more PopIII stellar generations,
namely massive and very massive stars of primordial chemical
composition. To do that we have adopted a recent version of the
two-infall chemical evolution model of CRM03 which reproduces the
majority of the observed features of the solar neighbourhood and the
whole Galaxy. By means of this model we have calculated the evolution
of several chemical elements in the Galactic interstellar medium,
focusing our attention on C, N, $\alpha$-elements (O, Mg, Si) and Fe,
by adopting several sets of stellar yields calculated for
zero-metallicity massive and very massive stars (PCSNe).  In
particular we have adopted the nucleosynthesis prescriptions from
Ober et al. (1983), Heger \& Woosley (2002), Umeda \& Nomoto (2002)
and Chieffi \& Limongi (2002, 2004). 

We have also tested some suggestions which appeared in the past years
in the literature concerning a variable IMF favoring massive stars in
the earliest phases of the evolution of the Galaxy. In particular, we
adopted the functional forms of the IMF as suggested by Larson (1998) and by
Hernandez \& Ferrara (2001). 

Our main results can be summarized as follows:
\begin{itemize}

\item[-] For the suggested values of the threshold metallicity for the
  formation of very massive PopIII objects, $Z_{\mbox{PopIII}}=
(10^{-6}-10^{-4}) \cdot Z_{\odot}$ (Schneider et al. 2002), the
effects of PopIII stars on the predicted abundances and abundance
  ratios are negligible even in the earliest Galactic evolutionary 
  phases, so that we cannot exclude their existence. Only if we assume 
that the PopIII regime lasted until $Z_{\mbox{PopIII}}=0.1Z_{\odot}$,
do we find noticeable differences in the predicted [X/Fe] vs. [Fe/H]
  trends and the agreement with the observations worsens relative to
  the standard case without PopIII.

\item[-] For the specific elements C and N, we find that in some cases
  (data relative to C/O from Ackerman et al., 2004) the  yields from
  PopIII can produce a better agreement with the
  observations. However, in this case, the agreement for all the other
  elements is worse relative to the standard model. The problem of
  N has been discussed and in particular the fact that new data at low
  metallicity seem to suggest a primary origin for this element
  produced in massive stars. We have shown that rather than invoking N
  production from PopIII stars, it is necessary to assume a continuous
  primary production from massive stars at all metallicities. Such a
  production still needs to be understood, although recent stellar
  models with rotation (Meynet \& Maeder, 2002) predict some primary
  production of N in massive stars, although not enough to explain the
  data (see Chiappini et al., 2005 for an extensive discussion on this
  point). 
\item[-]We did not discuss the effects of PopIII star nucleosynthesis
  on the Fe-peak elements (Co, Cr, Ni, Mn) and Zn, whose abundances in
  very metal poor stars require  changes in the standard yields (see
  also Fran\c cois et al. 2004), since UN02 had already discussed this
  point and concluded that PopIII stars cannot solve the problem and
  that nucleosynthesis in high-energy  hypernovae seems more
  promising. However, their analysis was based only on the direct
  comparison between element production ratios and abundances in very
  metal poor stars.   

\item[-] A continuously varying IMF, favoring massive stars in the
  early stages of Galactic evolution in the manner suggested by Larson
  (1998) does not produce acceptable results either for the abundance
  ratios or the G-dwarf metallicity distribution. Therefore, our
  conclusion is that a variable IMF in general produces results at
  variance with observations unless the variation is so small as to be
  irrelevant. The same conclusion was reached by Chiappini et
  al. (2000) by studying the evolution of the Galactic disk with the
  variable IMF proposed by Padoan et al. (1997). 

\item[-] We computed the number of PopIII SNe required to pollute the
  Galactic halo to a given $Z_{\mbox{PopIII}}$ and found that in order to
  pollute the ISM at the level of $Z_{\mbox{PopIII}}=10^{-4}Z_{\odot}$ we
  need a number of PCSNe varying from 2 to 20 according to the assumed
  stellar yields and that the time necessary to reach this situation
  is always of the order of few Myrs (from 1 to 5). This may
  explain why we do not see stars with zero-metallicity in the
  Galactic halo, simply because the number of these stars must have
  been negligible, even though it cannot be excluded that
  primordial stars formed in overdense regions and might have ended up
  in the Galactic bulge (where observations are more challenging)
  instead than in the halo. In any case, even the standard model of CRM03
  with a normal IMF including low mass stars predicts a fraction of
  stars with [Fe/H] $< -5.0$ of the order of $10^{-5}$, in agreement with
  current estimates. 
\end{itemize}

\subsection*{Acknowledgements}
F.M. and C.C. acknowledge financial support from the Italian MIUR (Ministry for
University and Scientific Research) through COFIN 2003, prot. 2003028039.
We also thank an anonymous referee for her/his helpful comments.


\begin{thebibliography}{}

\bibitem{00} Abel, T., Bryan, G.N., Norman, M.L., 2002, Science 295, 93

\bibitem {02} Akerman, C.J., Carigi, L., Nissen, P.E., Pettini, M.,
  Asplund, M., 2004, A\&A 414, 931

\bibitem{03} Asplund, M., Grevesse, N., Sauval, A.J., 2005, to
  appear in ``Cosmic abundances as records of stellar evolution and
  nucleosynthesis'', eds.  F.N. Bash  \& T.G Barnes, ASP Conference
  Series vol. XXX  (\emph{astro-ph/0410214})

\bibitem{04} Bromm, V., Larson, R.B., 2004, ARA\&A 42, 79,

\bibitem{06} Cayrel, R., Depagne, E., Spite, M., Hill, V., Spite, F.,
Fran\c cois, P., Plez, B., Beers, T., Primas, F., Andersen, J.,
Barbuy, B., Bonifacio, P., Molaro, P., Nordstr\"om, B., 2004, A\&A 416, 1117

\bibitem{07} Cen, R., 2003, ApJ 591, L5

\bibitem{08} Chabrier, G., 2003, PASP 115, 763

\bibitem{09} Chiappini, C., Matteucci, F., Ballero, S.K., 2005,
  A\&A 437, 429

\bibitem{10} Chiappini, C., Matteucci, F., Beers, T.C., Nomoto, K.,
  1999, ApJ 515, 226

\bibitem{12} Chiappini, C., Matteucci, F., Gratton, R., 1997, ApJ 477, 765

\bibitem{13} Chiappini, C., Matteucci, F., Meynet, G., 2003b, A\&A
  410, 257

\bibitem{14} Chiappini, C., Matteucci, F., Padoan, P., 2000, ApJ
  554, 1044

\bibitem{14a} Chiappini, C., Romano, D., Matteucci, F. 2003a, MNRAS
  339, 63 (CRM03)

\bibitem {15}Chieffi, A., Limongi, M., 2002, ApJ 577, 281

\bibitem {16}Chieffi, A., Limongi, M., 2004, ApJ 608, 405 


\bibitem{20} D'Antona, F., 1982, A\&A 115L, 1

\bibitem{22} Dickey, J.M., 1993, ed. R.M. Humphreys, ASP Conference
  Series 39, 93

\bibitem{24} El Eid, M.F., Fricke, K.J., Ober, W.W., 1983, A\&A 119, 54

\bibitem{26} Fran\c cois, P., Matteucci, F., Cayrel, R., Spite, M.,
  Spite, F., Chiappini, C., 2004, A\&A 421, 613

\bibitem{} Frebel, A., Aoki, W., Hiroyasu, A., Asplund, M., Barklem,
  P.S., Beers, T.C., Eriksson, K., Fechner, C., et al.  2005, Nature
  434, 871 

\bibitem{28} Gilmore, G., Wyse, R.F.G., Kuijken, K., 1989, ARA\&A 27, 555

\bibitem{30} Haiman, Z., Abel, T., Rees, M.J., 2000, ApJ 534, 11

\bibitem {32}Heger, A., Woosley, S.E., 2002, ApJ 567, 532

\bibitem{32a} Heger, A., Woosley, S.E., Waters, R., 2000, in ``The
  First Stars'', proceedings of the MPA/ESO workshop, A. Weiss,
  T.G. Abel, V. Hill (eds.), ESO Astrophysics Symposia, Springer,
  p. 93

\bibitem{33} Hernandez X., Ferrara A., 2001, MNRAS 324, 484

\bibitem{34} Israelian, G., Ecuvillon, A., Rebolo, R., Garcia-Lopez,
  R., Bonifacio, P. \& Molaro, P. 2004, to appear in A\&A
  (\emph{astro-ph/0405049})  

\bibitem{36} Iwamoto, K., Brachwitz, F., Nomoto, K., Kishimoto, N.,
  Umeda, H., Hix, W. Raphael, Thielemann, F.K., 1999, ApJS 125, 439

\bibitem{38} Jos\'e, J., Hernanz, M., 1998, ApJ 494, 680

\bibitem{40} Kennicutt, R.C. Jr., 1998, ApJ 498, 541

\bibitem{41} Larson, R.B., 1998, MNRAS 301, 569

\bibitem{42} Marigo, P., Chiosi, C., Kudritzki, R.P., 2003, A\&A 399,
  617

\bibitem{43} Matteucci, F., 1986, MNRAS 221, 911

\bibitem{44} Meynet, G., Maeder, A., 2002, A\&A 390, 561

\bibitem{45} Meynet, G, Maeder, A., Ekstr\"om, S., 2005, in ``The Fate
  of the Most Massive Stars'', ASP Conference Series 332, 232

\bibitem{46} M\'era, D., Chabrier, G., Schaeffer, R., 1998, A\&A
  330, 937

\bibitem{48}Nomoto, K., Thielemann, F.K., Yokoi, K., 1984, ApJ 286, 644

\bibitem{49}Ober, W.W.; El Eid, M.F., Fricke, K.J., 1982, in
  ``Supernovae: a survey of current research'', proceedings of the
  Advanced Study Institute, Cambridge, England, D. Reidel Publishing
  Co., p. 293

\bibitem{50}Ober, W.W., El Eid, M.F., Fricke, K.J., 1983, A\&A 119, 61

\bibitem{51}Oey, M.S., 2003, MNRAS 339, 849

\bibitem {52}Padoan, P., Nordlund, \AA., Jones, B.J.T., 1997, MNRAS
  288, 145

\bibitem{53} Portinari, L., Chiosi, C., Bressan, A., 1998, A\&A 334, 505

\bibitem{54} Rakavy, G., Shaviv, G., 1967, ApJ 148, 803

\bibitem{55} Rana, N., 1991, ARA\&A 29, 129

\bibitem{56} Renzini, A., Voli, M., 1981, A\&A 94, 175

\bibitem{57} Rocha-Pinto, H.J., Maciel, W.J., 1996, MNRAS 279, 447

\bibitem{58} Romano, D., Matteucci, F., 2003, MNRAS 342, 185

\bibitem{59} Sackett, P.D., 1997, ApJ 483, 103

\bibitem{60}Scalo, J., 1986, FCPh 11, 1

\bibitem {62} Schaye, J., Aguirre, A., Kim, T.S., Theuns, T., Rauch,
  M., Sargent, W.L.W., 2003, ApJ 318, 32

\bibitem{64} Schneider, R., Ferrara, A., Natarajan, P., Omukai, K.,
  ApJ 571,~30

\bibitem {66}Siess, L., Livio, M., Lattanzio, J., 2002, ApJ 570, 329

\bibitem{67}Sommer-Larsen, J., G\"otz, M., Portinari, L., 2003, ApJ
  596, 47

\bibitem{68} Spite, M., Cayrel, R., Plez, B., Hill, V., Spite, F.,
  Depagne, E., Fran\c cois, P., Bonifacio, P., Barbuy, B., Beers, T.,
  Andersen, J., Molaro, P., Nordstr\"om, B., Primas, F., 2005, A\&A, 430, 655


\bibitem{70} Tammann, G.A., Loeffler, W., Schroeder, A., 1994, ApJS 92, 487


\bibitem {72}Umeda, H., Nomoto, K., 2002, ApJ 565, 385

\bibitem {73} Umeda, H., Nomoto, K., 2005, ApJ 619, 427
\bibitem{74} Van den Hoek, L.B., Groenewegen, M.A.T., 1997, A\&AS 123, 305

\bibitem {76}Woosley, S.E., Weaver, T.A., 1995, ApJS 101, 181 

\bibitem{78} Wyse, R.F.G., Gilmore, G., 1995, AJ 110, 2771

\end{thebibliography}
\end{document}